\documentclass[journal]{IEEEtran}

\usepackage{amsmath}
\usepackage{amsfonts}
\usepackage{braket}
\usepackage{graphicx}
\usepackage{subcaption}
\usepackage{hyperref}
\usepackage[capitalise]{cleveref}
\usepackage{multirow}
\usepackage{multicol}
\usepackage{colortbl}
\renewcommand\vec{\mathbf}

\title{Renormalization group decoder for a four-dimensional toric code}
\author{
    \IEEEauthorblockN{K. Duivenvoorden\IEEEauthorrefmark{1}, N.P. Breuckmann\IEEEauthorrefmark{1}, B. M. Terhal\IEEEauthorrefmark{1}\IEEEauthorrefmark{2}\IEEEauthorrefmark{3}}\\
    \IEEEauthorblockA{\IEEEauthorrefmark{1}JARA Institute for Quantum Information, RWTH Aachen University, 52056 Aachen\\
    \{kasperd, breuckmann\}@physik.rwth-aachen.de}\\
    \IEEEauthorblockA{\IEEEauthorrefmark{2}Forschungszentrum J\"ulich GmbH, J\"ulich, Germany}\\
    \IEEEauthorblockA{\IEEEauthorrefmark{3}QuTech, Delft University of Technology, Lorentzweg 1, 2628 CJ Delft, The Netherlands\\
    \{bterhal\}@gmail.com}
}

\begin{document}

\maketitle

\begin{abstract}
We describe a computationally-efficient heuristic algorithm based on a renormalization-group procedure which aims at solving the problem of finding minimal surface given its boundary (curve) in any hypercubic lattice of dimension $D>2$. We use this algorithm to correct errors occurring in a four-dimensional variant of the toric code, having open as opposed to periodic boundaries. For a phenomenological error model which includes measurement errors we use a five-dimensional version of our algorithm, achieving a threshold of $4.35\pm0.1\%$. For this error model, this is the highest known threshold of any topological code. Without measurement errors, a four-dimensional version of our algorithm can be used and we find a threshold of $7.3\pm0.1\%$. For the gate-based depolarizing error model we find a threshold of $0.31\pm0.01\%$ which is below the threshold found for the two-dimensional toric code.
\end{abstract}


\section{Introduction}
The threshold of an error correcting code family is an indicator of whether the code can be used for a quantum memory. This is because for any experimental QC platform so far it is challenging to obtain high coherence qubits with two-qubit gate fidelities below $10^{-3}$.
Topology has been a key ingredient in finding new codes with thresholds higher or comparable to those obtained via concatenation~\cite{Knill}, see~\cite{terhal:review} and references therein. The prime example is the surface code, with a threshold as high as $1.1\%$~\cite{WFHL:gatebased}. Two or three-dimensional color codes are also a promising family of topological codes due to the ability to perform the Clifford gates transversally (2D color codes) or even $T$ gates transversally (3D color codes)~\cite{CTV:review}. For two-dimensional color codes decoding methods have led to thresholds of $0.082\%$~\cite{LAR:colorcode} (4.8.8 color code) and $0.3\%$~\cite{thesis:beverland} (6.6.6 color code). These numbers are lower than the surface code which can partially be accounted for by the weight of the stabilizers, being maximally 8 or 6, respectively. 
In general, these thresholds are only indications of what to expect experimentally, since real errors can be biased, stochastic or coherent, non-Pauli, leaky or induce cross-talk and one can expect further code optimizations which are platform-dependent. 

Stability of a quantum memory is often related to the dimensionality of the support of logical operators. Excitations in the toric code are point-like anyons which can diffuse over large distances without energy cost, leading to thermal instability~\cite{AFH:2Dthermal}. The situation improves for the three-dimensional Haah code, where logical operators are no longer one-dimensional. However,  for this code the memory time increases with increasing system size only up to a certain critical size~\cite{BH:3Dthermal} after which entropic factors shorten the memory time. 
Going up yet another dimension, the four-dimensional toric code~\cite{DKLP}, having logical operators with two-dimensional support, is stable under thermal fluctuations below a certain non-zero critical temperature~\cite{AHHH:4Dthermal}. In error correcting terms, the stabilizer checks have a local linear dependency which can be used to repair the erroneous syndrome data~\cite{bombin:selfcorrection}, obviating the need for making the syndrome record reliable by repetition in time. Such a single-shot correction schedule can also be used in decoding a three-dimensional gauge color code~\cite{Bombin:singleshot, BNB:3dcolorcode}. 
In this paper we will study whether the four-dimensional toric code can have a higher threshold than the surface code, despite having higher weight stabilizers.

We will discuss a version of the four-dimensional toric code having open boundary conditions, which we will call the \textit{tesseract code}. Its construction is analogous to the construction of the surface code and can be formalized using relative homology~\cite{BK:surfacecode,FM:surfacecode}. The tesseract code encodes a single qubit instead of the six logical qubits encoded in the four-dimensional toric code. The main reason to study the tesseract code, as opposed to the four-dimensional toric code, is that one can globally correct faulty syndrome data before decoding, giving rise to a single-shot ``repair syndrome'' decoder.

Of course, a four-dimensional code is not appealing to be implemented in purely 2D chip-based architecture. However, modular-based architectures such as nitrogen vacancy centers in diamond~\cite{CH:NVcenters} or networks with few qubit nodes~\cite{Ret:Network} could potentially have less strenuous constraints on connectivity. Clearly, embedding a 4D code into 3D space requires a long-range connectivity between qubits which grows with the size of the code. In practice one realizes a code of finite size and even expander-like graphs can be embedded in 3D physical space when time-multiplexing of connections can be used, e.g. the internet.

The parameters of the tesseract code family are $[[6L^4-12L^3+10L^2-4L+1, 1,L^2]]$ (see Section~\ref{sec:defcode}) as compared to $[[L^2+(L-1)^2,1,L]]$ for the regular 2D surface code~\cite{DKLP}, implying that one can have a $[[33,1,4]]$ tesseract code versus a $[[25,1,4]]$ surface code, or a $[[241,1,9]]$ tesseract code versus a $[[145,1,9]]$ surface code. All checks in the bulk of the tesseract code act on 6 qubits while each qubit participates in 8 different checks (qubit degree 8). Table~\ref{tab:smallcodes} in Appendix~\ref{A:smallcodes} presents several small codes which interpolate between the surface code and the tesseract code. In Table~\ref{tab:thresholds} we summarize the known thresholds and the new results obtained with the RG decoder. In Appendix~\ref{sec:logic} we briefly comment on what is known on getting universal logic using a 4D toric code.

\begin{table} 
\centering
 \begin{tabular}{lc||cc|cc}
 \multirow{3}{*}{code}&dimension &\multicolumn{4}{c}{measurement}\\
 &of support of &\multicolumn{2}{c|}{perfect}&\multicolumn{2}{c}{faulty}\\
& logical operator & & \\ 
\hline
 1D Ising &1& \multicolumn{2}{c|}{$\cellcolor[gray]{0.9}50.0\%$} &\multicolumn{2}{c}{$\cellcolor[gray]{0.9}11.0\%$} \\
 2D surface &1& \multicolumn{2}{c|}{$\cellcolor[gray]{0.9}11.0\%$ \cite{WHP:threshold}} &\multicolumn{2}{c}{$\cellcolor[gray]{0.9}3.3\%$ \cite{WHP:threshold}} \\
 3D cubic &1&\multicolumn{2}{c|}{$\cellcolor[gray]{0.9} 3.3\%$}& \multicolumn{2}{c}{?}  \\ 
 \hline
 2D Ising &2& \multicolumn{2}{c|}{$\cellcolor[gray]{0.9}50\%$}&\multicolumn{2}{c}{$17.2\%$}\\
 3D cubic &2& \multicolumn{2}{c|}{$17.2\%$}&$7.3\%$& $\cellcolor[gray]{0.9}11.0\%$ \\
 4D tesseract&2& $7.3\%$&  $\cellcolor[gray]{0.9}11.0\%$ \cite{takeda}&\multicolumn{2}{c}{$4.35\%$ }
 \end{tabular}
 \caption{Overview of thresholds for surface codes of different dimensions, using a phenomenological error model with perfect or faulty syndrome measurement, as explained in Section~\ref{sec:phen}. Error correction for these codes proceeds independently for $X$- and $Z$-errors and protection from logical $\overline{X}$ and $\overline{Z}$ errors depends on the dimensionality of the support of the logical operator. For the 1D and 2D Ising model, which essentially represent classical codes, we only list the data point for the logical operator with extensive support, its logical partner has $0$-dimensional support and hence no threshold exists. The gray values are previously determined upper bounds while all other values are new lower bounds on the threshold obtained using the efficient RG decoder introduced in this paper. In Section~\ref{sec:phen} and Appendix~\ref{A:HD} we explicitly show how in our error model the decoding problem of a $D$-dimensional code with perfect syndrome measurement is equivalent to space-time decoding of a $D-1$-dimensional code with faulty syndromes, leading to the same thresholds as shown in the Table.}
 \label{tab:thresholds}
\end{table}

Earlier thresholds of the four-dimensional toric code have been found to be as high as $1.59\%$ for the phenomenological error model with faulty syndrome measurements \cite{BDMT:localdecoders}. These decoders attempted to minimize the local curvature of syndrome curves in order to shrink these curves or apply a 4D version of Toom's rule. It was observed in \cite{BDMT:localdecoders} that the limiting factor of the decoder was the occurrence of stripes of errors, having a width larger than the decoding region. Such stripes have straight syndrome curves as boundaries, with no curvature. Hence the corresponding errors would not be corrected by the decoder.

We will introduce a new decoder for the tesseract code based on a renormalization scheme. Renormalization group (RG) decoders have been successfully used before to decode the toric code \cite{DP:fast,DP:RG}, the qudit toric code \cite{ABCB:RGqudit}, color codes \cite{SR:RG_colorcode} and the Haah code \cite{BH:3Dthermal}. This class of decoders can be divided into two groups: the soft-decision RG decoders function by renormalizing the lattice whereas hard-decision decoders function by renormalizing the scale at which syndromes are clustered. 
Our decoder falls in the first group: we describe a way of coarse-graining the four-dimensional lattice of the tesseract code. Then, the aforementioned stripes are no longer a limiting factor of the decoder due to the doubling of the decoding region at each RG step. It is also possible to use a hard-decision RG decoder for the tesseract code as the efficient decoder in \cite{BH:3Dthermal} works for any topological code. However, the Bravyi-Haah RG procedure is not fine-tuned to the decoding problem at hand, namely finding a minimal surface given its boundary, and we expect it to be non-optimal. 
We report on thresholds of our decoder using both the phenomenological error model as well as a gate- or circuit-based error model in order to objectively compare with other codes. 

Our paper is organized as follows. In the Section~\ref{sec:code} we will introduce the tesseract code. We will explain how to view this code from the perspective of relative homology and discuss why it encodes one qubit. 
In Section~\ref{sec:noise} we explain that when using a phenomenological error model, minimum-weight space-time decoding translates to finding a surface of minimal area having a given curve as its boundary: in Appendix~\ref{A:HD} we show how this holds generally for high-dimensional surface codes, basically following the line of thinking introduced in \cite{DKLP}.
In Section~\ref{sec:gb} we describe the gate-based error model in detail. In Section~\ref{sec:decoder} we will explain the RG decoder. In Section~\ref{sec:results} we report on the numerical results of a single-shot decoder and the RG decoder. We end the paper with some concluding remarks in Section~\ref{sec:con}.

\section{The code}
\label{sec:code}
        The tesseract code can be understood on various levels of abstraction. The most straightforward way to define the code is to introduce sets of edges, faces and cubes and associate qubits with faces and $X$- and $Z$-stabilizers with edges and cubes respectively. We will refer to the low-weight (not necessarily independent) generators of the stabilizer group as stabilizers or check operators. In Sections~\ref{sec:hom_des} and~\ref{sec:cel_com} we will be a bit more formal and review the concept of homological CSS codes based on cellular complexes and show how the tesseract code can be viewed as an example of such code using relative homology. In Section~\ref{sec:cel_com} we will argue that the tesseract code encodes 1 qubit, using a deformation retraction argument. For the less formally-inclined Sections \ref{sec:hom_des} and \ref{sec:cel_com} can be skipped.

\subsection{Definition}
\label{sec:defcode}
We start by defining cells ($o$). or more specifically, cells can be edges ($e$), faces ($f$) or cubes ($c$). Let $\vec a_i $ for $i\in \{1,2,3,4\}$ denote four unit-length basis vectors of $\mathbb{R}^4$. We will consider cells which are oriented along these four directions, i.e.
\begin{align}\label{eq:edges}
 e_{\{i\}}(\vec v)    &:= \{\vec v + s \vec a_i \ \ |\ \ s\in[0,1]\} \ \ , \\ \label{eq:faces}
 f_{\{i,j\}}(\vec v)    &:= \{\vec v + s_1 \vec a_i +s_2\vec a_j\ \ |\ \ s_1, s_2 \in[0,1]\} \ \ , \\ \label{eq:cubes}
 c_{\{i,j,k\}}(\vec v)   &:= \{\vec v + s_1 \vec a_i +s_2 \vec a_j+s_3 \vec a_k\ \ |\ \ s_1,s_2, s_3 \in[0,1]\} \ \ ,
\end{align}
where the vector $\vec v = \sum_i v_i\vec a _i$ has integer coordinates $v_i$. Consider the spaces $B\subset U\subset \mathbb{R}^4$ being $U = [0,L_1-1]\times [0,L_2-1]\times [0,L_3]\times [0,L_4]$ and $B$ the union of four hyperplanes defined by the restriction $v_3\in \{0,L_3\}$ and  $v_4\in \{0,L_4\}$. We will generally set all lengths $L_i$ equal to $L$. Alternatively, one can set some lengths to one to obtain lower-dimensional versions of the tesseract code, see Table~\ref{tab:smallcodes} in Appendix \ref{A:smallcodes}. 

A cell $o$ is said to be contained in a space, say $U$, when $o\subset U$. The face set, defined as $F_L$, consists of those faces contained in $U$ but not contained in $B$. Similarly, the edge set $E_L$ and cube set $C_L$ consist of those edges and cubes, respectively, which are contained in $U$, but not contained in $B$. The cardinality of these sets are given by (see Appendix~\ref{A:counting}):
\begin{align} \nonumber
|C_L| = |E_L| &= 4L^4-8L^3+6L^2-2L \ \ ,\\ \nonumber
|F_L| &= 6L^4-12L^3+10L^2-4L+1 \ \ .
\end{align}

\begin{figure}[hbt]
\centering
 \includegraphics[scale = .5]{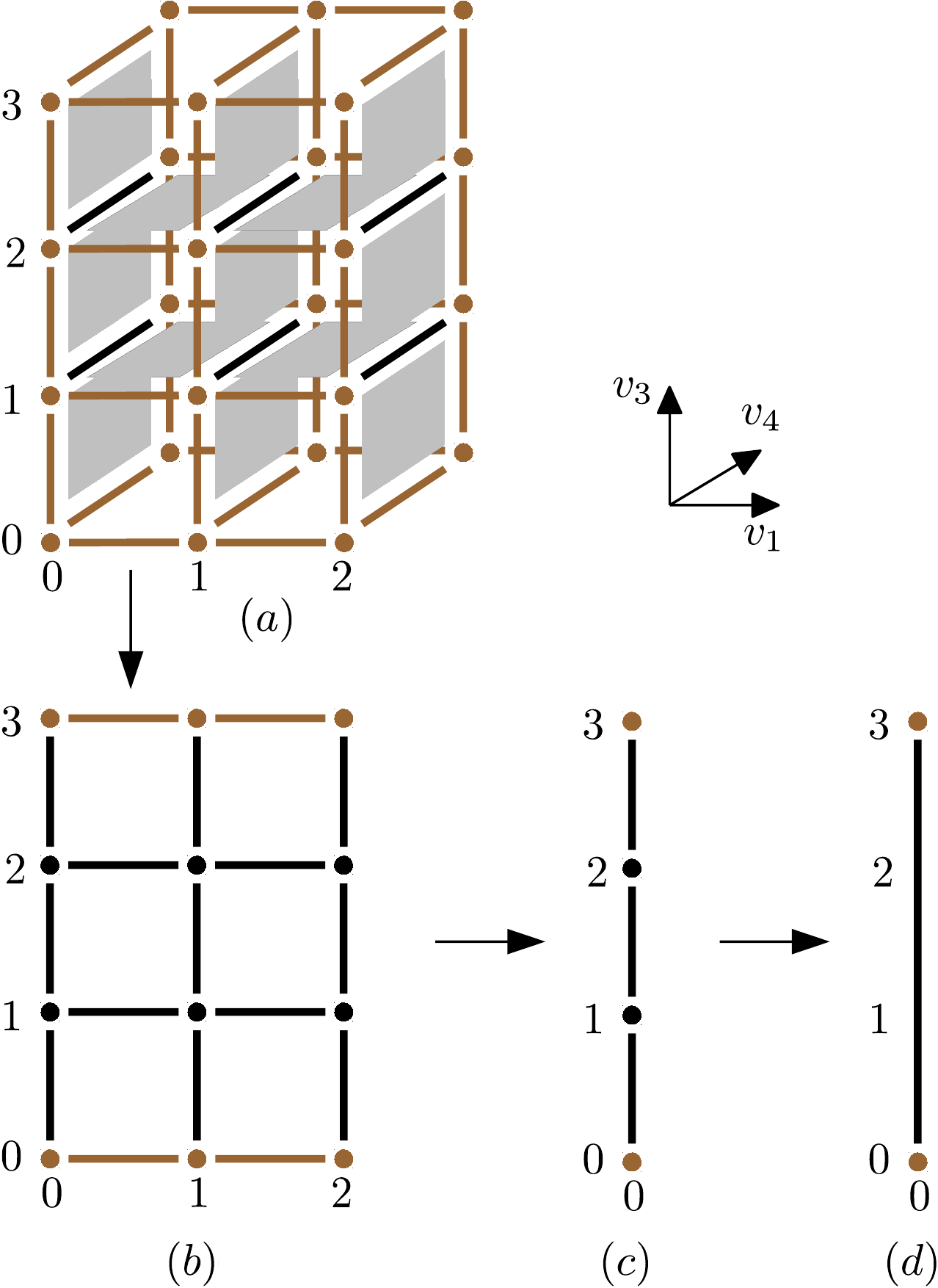}
 \caption{\label{fig:surfacecode}(color online) Panel (a): $v_2 = 0$ cross-section of the cellular complex corresponding to the surface code of distance 3. Brown points and lines indicate vertices and edges which are contained in $U$ but also in $B$. Black lines and gray squares indicate edges and faces which are only contained in $U$. Each such face contains a qubit. Cubes are not indicated. Panel (b): same cellular complex with points indicating edges (oriented in the $\vec a_4$ direction) and lines indicate faces. Points and lines in brown are again contained in $U$ but also in $B$. Panel (c): graphical representation of projecting out the $\vec a_1$ direction coordinate by means of a deformation retraction. Panel (d): simplified cellular complex representing only 1 qubit.}
\end{figure}

Having constructed the sets $E_L$, $F_L$ and $C_L$, we can straightforwardly define the tesseract code of size $L$. Qubits are defined for each face in $F_L$. The $X$- and the $Z$-stabilizers of the code are defined for each edge in $E_L$ and each cube in $C_L$ respectively. Their action is determined by the inclusion, $e \subset f \subset c$ of edges, faces and cubes:
\begin{align}\label{eq:sx}
 S^{X}_e := \prod_{f:e\subset f} X_f \ \ , \\ \label{eq:sz} 
 S^{Z}_c := \prod_{f:f\subset c} Z_f\ \ .
\end{align}
Both $X$- and $Z$-stabilizers are maximally of weight six (act on 6 qubits non-trivially). Note how the tesseract code is a higher-dimensional version of the surface code. The surface code is obtained by setting $L_2=L_4 = 1$ and $L_1=L_3=L$. See Fig.~\ref{fig:surfacecode}(a) for an illustration of a distance-3 surface code. In this case only those edges $e_{\{i\}}(\vec v)$ are contained in $E_L$ for which $i$ is four. Hence they effectively reduce to vertices when ignoring this dimension. Similarly, faces on which the qubits live, reduce to edges and cubes reduce to faces. Setting only $L_2=1$ and all other lengths equal to $L$ one obtains a three-dimensional code, which we will refer to as the cubic code (not be confused with the Haah code \cite{BH:3Dthermal}).

The tesseract code encodes a single qubit, as we will argue in Section~\ref{sec:cel_com}. Here we simply give representatives of the logical operators:
\begin{align} 
 \overline{X} &= \prod_{v_1,v_2=0}^{L-1} X_{f_{\{3,4\}}(v_1\vec a_1 +v_2\vec a_2 )} 
\label{eq:logx} \ \ \text{and}\\ 
 \overline{Z} &= \prod_{v_3,v_4=0}^{L-1} Z_{f_{\{3,4\}}(v_3\vec a_3 +v_4 \vec a_4 )} \ \ .
\label{eq:logz}
\end{align}
These operators anti-commute since they overlap on a single qubit, corresponding to the face $f_{\{3,4\}}(\vec 0)$. Comparing the tesseract code with the surface code gives insight into how the space $B$ changes the boundary conditions. In the surface code, $B$ ensures that two of the four boundaries are `rough', see Fig.~\ref{fig:surfacecode}(b), so that $\overline{Z}$ can begin and end at this rough boundary (meaning that it commutes with the $X$-stabilizers). Similarly, on a smooth boundary $\overline{X}$ can begin or end. If $B$ were the empty set, all boundaries would be `smooth' and no qubit would be encoded since any $\overline{X}$ could be deformed and contracted to a $X$-stabilizer. 

For the tesseract code all boundaries associated with the third and fourth direction are `rough' (i.e. these boundaries are formed by setting the third and fourth coordinate to their boundary values) whereas boundaries associated to the first and second direction are `smooth'. Thus the logical $\overline{Z}$ in Eq.~(\ref{eq:logz}) which fully lies in the plane spanned by the vectors $\vec a_3$ and $\vec a_4$ is a surface attached at rough boundaries, while the logical $\overline{X}$ is a surface attached at smooth boundaries.

Summarizing the tesseract code encodes a single qubit using $6L^4-12L^3+10L^2-4L+1$ physical qubits for a distance $L^2$. In Appendix~\ref{A:HD} we generalize the surface code family, which includes the surface and tesseract code, to a general $(d_1,d_2)$-surface code having a $d_1$-dimensional $\overline{X}$ operator and a $d_2$-dimensional $\overline{Z}$ operator. In Appendix~\ref{sec:logic} we argue that one can only perform those logicals by a constant depth circuit, which are element of the restricted Pauli group.

\subsection{Homological description}
\label{sec:hom_des}
The tesseract code is a homological CSS code \cite{BK:surfacecode, FM:surfacecode} in the sense that stabilizers can be defined in terms of boundary operators. Let $C_i$ (for $i\in\{1,2,3\}$) be vector spaces over $\mathbb{F}_2$. In the next section we describe how one can obtain these vector spaces from a cellular complex, using (relative) homology, here we state and use their properties to define the tesseract code. Elements of $C_1$ are formal sums of edges ($C_1 \ni \sum_e E_e e$, $E_e\in\mathbb{F}_2$), and similarly, elements of $C_2$ and $C_3$ are formal sums of faces and cubes respectively. An element of $C_k$ is also referred to as a $k$-chain. The different spaces $C_k$ are related by boundary operators $\partial_k$:
\begin{align}\label{eq:cc}
  C_3 \xrightarrow{\partial_3} C_2   \xrightarrow{\partial_2} C_1 \ \ .
\end{align}
They can be most easily defined by specifying their action on the basis vectors of $C_2$ and $C_3$, that is $\partial(f) = \sum_{e\subset f}e $, the sum of the  (up to) four edges of a face $f$, and $\partial(c) = \sum_{f\subset c}f$, the sum of the (up to) six faces of a cube $c$. The transpose of the boundary operator is the co-boundary operator: $\delta_{k} := \partial_{k+1}^T$. Pauli operators can be labeled by a pair of face sets $a,b\in C_2$, i.e. $P_{a,b}:=\prod_fX_f^{\alpha_{f}}\prod_fZ_f^{\beta_{f}}$, where $a = \sum_f \alpha_f f$ and $b=\sum_f \beta_f f$. One has $P_{a,0}P_{0,b} = (-1)^{\braket{a,b}}P_{0,b}P_{a,0}$ with $\braket{a,b}:=\sum_f \alpha_f \beta_f \in \mathbb{F}_2$. Stabilizer generators are given by applying $\delta_1$ and $\partial_3$ on basis vectors in $C_1$ and $C_3$ respectively, i.e. $S^{X}_e = P_{\delta_1(e),0}$ and $S^{Z}_c = P_{0,\partial_3(c)}$.
Their commutation follows from $\partial_2\circ\partial_3=0$ via $\braket{\delta_1(e),\partial_3(c)} = \braket{e,\partial_2\circ \partial_3(c)}$. Logical $\overline{Z}$ operators $P_{0,z}$ should satisfy $\partial_2 (z) = 0 $ in order to commute with all $X$-stabilizers. Also, $P_{0,z}$ can be written as a product of stabilizer generators if $z$ is in the image of $\partial_3$. Hence $\text{dim}(H_2)$, where $H_2 = \frac{\text{Ker}(\partial_2) }{\text{Im}(\partial_3)}$ equals the number of logical qubits\footnote{Formally, the homology group $H_i\equiv H_i(\mathbb{Z}_2)$ since we have set $G=\mathbb{Z}_2$, i.e. the addition group of $\mathbb{F}_2$ and homology groups can be defined over general groups $G$.}.

\subsection{Cellular complex}
\label{sec:cel_com}
A more precise description of the tesseract code is in terms of cellular complexes.  The building blocks of cellular complexes are $k$-cells, which are spaces isomorphic to an $k$-dimensional closed ball. Vertices are $0$-cells, edges are 1-cells, faces are 2-cells etc. We will refer to 4-cells as hyper-cubes and they can be defined analogous to the lower-dimensional cells in \cref{eq:edges,eq:faces,eq:cubes}:  $h(\vec v)  := \vec v + [0,1]^4$.

By definition, the boundaries of cells are part of the cellular complex. For our specific complex, it is clear what the boundaries of cells are. For example, the boundary of the edge $e_{i}(\vec v )$ is simply the union of the two vertices $\vec v$ and $\vec v+\vec a_i$, the boundary of a face $f$ is the union of four edges, etc. Note that not all boundaries of faces in $F_L$ are contained in the edge set $E_L$ since $E_L$ does not contain edges fully contained in $B$. However, by construction $U$ and $B$ both form cellular complexes. A cellular complex $U$ naturally comes with vector spaces $C_k(U)$ over $\mathbb{F}_2$, which are formal sums of $k$-cells in $U$, i.e. $k$-chains, and a boundary map $\partial_k$ between these spaces, specifying the boundary of $k$-cells:
\begin{align}\nonumber
  0 \xrightarrow{\partial_5} C_4(U) \xrightarrow{\partial_4} C_3(U)  &\xrightarrow{\partial_3}  C_2(U) \xrightarrow{\partial_2}  \\ \nonumber
    & C_1(U)  \xrightarrow{\partial_1}  C_0(U) \xrightarrow{\partial_0} 0 \ \ .
\end{align}
These maps satisfy $\text{Im}(\partial_{k+1})\subset \text{Ker}(\partial_k)$ or equivalently $\partial_{k}\circ \partial_{k+1} = 0$,  which generalizes the fact that boundaries of surfaces are always closed curves, to higher-dimensional chains. The last map $\partial_0\colon C_0(U) \rightarrow 0$ simply states that vertices have no boundaries and the first map $\partial_5 \colon 0\rightarrow C_4(U)$ states that hypercubes aren't boundaries of five-dimensional chains. The map $\partial_k$ can be restricted to act on the quotient space $C_k(U,B) := C_k(U)/C_k(B)$, in which $k$-chains in $C_k(U)$ which differ by $k$-chains in $C_k(B)$ are identified, i.e. one defines a quotient boundary map $\partial_k^B$ which maps from $C_k(U,B)$ to $C_{k-1}(U,B)$. Loosely speaking, the quotient procedure can be viewed as considering only formal sums of $k$-cells contained in $U$ and not $B$, that is, the maps $\partial_k^B$ (for $k\in\{2,3\}$) are equal to the maps given in Eq.~(\ref{eq:cc}). The quotient boundary map (and its associated quotient co-boundary map) have similar properties as the boundary map themselves, i.e. the boundary of the boundary is 0. The relevant objects are now the relative homology groups $H_k(U,B) = \frac{\text{Ker}(\partial_k^B) }{\text{Im}(\partial_{k+1}^B)}$. Specifically, $H_2(U,B)$ and $\dim(H_2(U,B))$ determine the logical operators and the number of logical qubits of the code when we use the construction described in the previous Section. What surface is homologically non-trivial, i.e. contained in $H_2(U,B)$, is now determined relative to the boundary $B$.  

So far we have argued how the tesseract code can be defined using the language of cellular relative homology. We will use this to argue that $\text{dim}(H_2)=1$, i.e. the tesseract code encodes 1 logical qubit. This may not be surprising, but it is useful to see how this follows from arguments which are more generally applicable to homological CSS codes. If we were to choose the space $U$ as the four-dimensional torus $T^4$, i.e. identify vertices at opposite boundaries in all directions, the corresponding code is the 4D toric code and the number of logical qubits equals $\dim(H_2(T^4))=6$. These 6 logical operators of the 4D toric code correspond to closed toric surfaces. 

The homology groups $H_k(U,B)$ are isomorphic to $H_k(U', B')$ where $B'$ and $U'$ are obtained via a so-called \textit{deformation retraction} from $B$ and $U$. A deformation retraction is the process of shrinking a topological space $X$ (within itself) into a subspace $Y\subset X$. It is a continuous map $f:[0,1]\times X \rightarrow X$ such that $f(0,X) = X$, $f(1,X) = Y$ and $f(t,\cdot)$ acts as identity on $Y$ for all $t$ \cite{book:hatcher}. In a first step we will simplify the quotient space $U/B$ using a specific deformation retraction. In a second step we will relate this simplified space to a code having only 1 qubit and no stabilizers.

Let $U' = \{0\}^2\times[0,L]^2 \subset U$ and let $B' = \{ v_3 \vec a_3 +v_4 \vec a_4 \;| \;v_3 \in\{0,L\} \vee \; v_4 \in\{0,L\}\} \subset B$ and note that $B'\subset U'$. The map  $f(t, \sum_i v_i\vec a _i) = (1-t) (v_1\vec a _1 + v_2\vec a _2) + v_3\vec a _3 +v_4\vec a _4$ is a deformation retraction of $U/B$ into $ U'/B'$ and hence $\dim(H_2(U,B))= \dim(H_2(U', B'))$. In order to calculate the dimension of the homology groups of $U'/B'$, we explicitly construct cellular complexes for these two spaces. The cells
\begin{align}\nonumber
 f' &= \{0\}^2\times[0,L]^2, \\\nonumber
 e'_1 &= \{0\}^2\times[0,L]\times \{0\}, \ \  e'_2 = \{0\}^2\times[0,L]\times \{L\}, \ \ \\ \nonumber
 e'_3 &= \{0\}^2\times\{0\}\times[0,L], \ \ e'_4 = \{0\}^2\times\{L\}\times[0,L], \\\nonumber
 v'_1 &= (0,0,0,0), \ \ v'_2 = (0,0,0,L), \ \ \\ \nonumber
 v'_3 &= (0,0,L,0), \ \ v'_4 = (0,0,L,L),
\end{align}
form a cellular complex of $U'$ and the cells $\{e'_i\}$ and $\{v'_i\}$ form a cellular complex of $B'$. Clearly, $f'$ is the only $2$-cell of $U'$ which is not contained in $B'$. The spaces $C_k(U')/ C_k(B')$ are all equal to zero except for $C_2(U')/ C_2(B') = \mathbb{F}_2$, hence all homology groups $H_k(U', B')$ are trivial except for $H_2(U', B')$ which is one-dimensional.
In error correcting terms, the corresponding code consists of a single qubit and no stabilizers and hence trivially encodes one qubit. 
Specifically, one has $\text{dim}(H_1) = 0$ or, in words: all closed curves in a tesseract code are the boundary of some surface. This is not true for the 4D toric code, i.e. $\text{dim}(H_1(T^4))=4$. This is an important difference between these codes and it allows us to study the single-shot decoder in Section~\ref{sec:ss} for the tesseract code. In this decoder the erroneous syndrome, which is a set of open curves with a zero-dimensional boundary is first repaired to be a set of closed curves. Since any set of closed curves are the boundary of some error surface in the tesseract code, the decoder can find a set of qubit errors.
A similar deformation retraction argument, graphically given in Fig.~\ref{fig:surfacecode}(b-d), can be used to show that the surface code encodes 1 qubit.

For completeness, we check that the number of logical qubits is consistent with the number of physical qubits and stabilizers. Let $V_L$ be the set of vertices in $U\backslash B$  having integer coefficients. \footnote{Backslash denotes set subtraction and $/$ denotes taking the quotient.} Its cardinality is given by $|V_L| = L^4-2L^3+L^2$, see Appendix~\ref{A:counting}. For each vertex $\vec v \in V_L$  there is a linear dependency between the $X$-stabilizers:
\begin{align}\label{eq:sx_lin_dep}
 \prod_{e:\vec v \in e} S^{X}_e =\mathbb{I} \ \ .
\end{align}
This is a consequence of $\delta_1\circ\delta_0 = 0$.  Assuming that all linear dependencies between $X$-stabilizers are of the form given in Eq.~\eqref{eq:sx_lin_dep} and that labeling them with a vertex $\vec v\in V_L$ does not lead to overcounting, the number of logical qubits is given by $|F_L| - 2(|E_L|-|V_L|) = 1$. 

A further comment on the tesseract code is this. In dimensions higher than two one has to be careful to distinguish objects with non-trivial topology (in the sense of having non-trivial homology groups) from the action of operators associated with these objects on the code space. For example, for the 4D toric or tesseract code, one can consider a Klein bottle error, that is, a non-orientable topologically non-trivial $Z$-error surface which can be embedded without intersection in four-dimensional space. The Klein bottle surface has no boundary and is thus not detected by any stabilizer. The Klein bottle error has in fact trivial effect on the code space since it can be constructed from stabilizer cube operators of the code. A Klein bottle error can also be represented in the cubic code (with the convention of placing qubits on faces) but since any such representation must be necessarily self-intersecting in a three-dimensional space, the error does not commute with all stabilizers of the code, nor can it be made from cube operators. In 3D it thus constitutes a genuine excitation out of the code space.

\section{Error models}
\label{sec:noise}
We assess the performance of the tesseract code by testing whether it can correct errors which are applied according to two different types of error models: phenomenological errors and gate-based errors. The main reason for using a phenomenological error model is that minimum-weight decoding has a straightforward geometrical interpretation. This error model however doesn't take into account that weight-six stabilizers are technically demanding to measure. The gate-based error model takes into account the full circuit for measuring the different stabilizers, that is, all elements in the circuit, including CNOT gates, ancilla creation and measurement and idling gates, are assumed to undergo depolarizing errors. 

\subsection{Phenomenological Error Model}
\label{sec:phen}
The phenomenological error model assigns errors to each qubit independently. Pauli operators are applied according to the following distribution $\mathbb{P}(\mathbb{I}) = (1-p)^2$, $\mathbb{P}(X) = p(1-p)$, $\mathbb{P}(Y) = p^2$ and $\mathbb{P}(Z) = p(1-p)$. Moreover, the measurement data is also assumed to be faulty, which is modeled by a bit-flip channel with parameter $q$. 
Due to the independence of Pauli $X$- and $Z$-errors, and since Pauli $Z$-errors only affect the outcome of $X$-stabilizers (and vice versa), the decoding problem separates into finding a Pauli $Z$-error plus $X$-stabilizer measurement errors which together are consistent with $X$-stabilizer measurements, and a similar set of qubit and measurement errors consistent with $Z$-stabilizer measurements. Moreover, these two problems are equivalent since they are mapped onto each other via duality of the lattice. It is hence sufficient to only discuss the decoding of Pauli $Z$-errors in combination with $X$-stabilizer measurement errors. 

Due to the independence of single qubit errors, it is appropriate to use a minimum-weight space-time decoding algorithm, by which we mean: given the outcome of repeated faulty measurement in time of all stabilizers, find the minimal number of qubit and measurement errors that could have led to this outcome. When $q=0$ and no measurements are repeated this reduces to minimum-weight decoding.

We will now discuss how, for the tesseract code, minimal-weight (space-time) decoding translates to finding a minimal surface having a given curve as its boundary in a 4+1 space-time cellular complex, in complete analogy with 2+1 space-time decoding for the surface code \cite{DKLP}. As a warm up, we will first do so for $q=0$, which is also explained in \cite{BDMT:localdecoders}. Let $f_{\text{error}}$ be the face set corresponding to the Pauli $Z$ error $P_{0,f_{\text{error}}}$. The syndrome $e_{\text{synd}}\in C_1$ is a formal sum of edges corresponding to $X$-stabilizers anti-commuting with the error, i.e. $e_{\text{synd}} = \sum_e \sigma_e e$ where
$(-1)^{\sigma_e}$ is the $\pm 1$ eigenvalue outcome of the stabilizer $S_e^X$. Note that the outcome of measurement of $S_e^X$ depends on the overlap of $f_{\text{error}}$ and $\delta_1 (e)$, i.e. $\sigma_e = \braket{f_{\text{error}},\delta_1(e)}$. Hence, the syndrome is exactly the boundary of the face set corresponding to the error due to:
\begin{align} \nonumber
 e_{\text{synd}} = \sum_e \braket{f_{\text{error}},\delta_1(e)} e  = \sum_e \braket{\partial_2(f_{\text{error}}), e} e = \partial_2(f_{\text{error}}) \ \ .
\end{align}
It follows that any valid Pauli $Z$ correction (i.e. giving rise to the same measurement outcome $e_{\text{synd}}$) is labeled by a face set $f_{\text{cor}}$ satisfying, $\partial_2(f_{\text{cor}}) =  e_{\text{synd}} $. Hence minimum-weight decoding translates to finding a minimal surface having a given curve as its boundary. This can be compared to decoding the surface code, where minimal-weight perfect matching finds strings of minimal length having a given set of vertices as its boundary.

Now consider faulty measurements with $q=p$. We define edges $\tilde e_{\{i\}}(\vec v)$, faces $\tilde f_{\{i,j\}}(\vec v)$ and cubes $\tilde c_{\{i,j,k\}}(\vec v)$ as subspaces of $\mathbb{R}^5$, analogous to Eqs.~\eqref{eq:edges}-\eqref{eq:cubes} with the difference that the directions $i$, $j$, and $k$ can also take the value 5. The space $\mathbb{R}^5$ is spanned by five basis vectors $\{\vec a_i\}_{i=1}^5$ and we will refer to the $\vec a_5$ direction as time. Again, we will only consider edges, faces and cubes which are contained in a space-time cellular complex $U_{\mathrm{ST}}$ and not contained in $B_{\mathrm{ST}} \subset U_{\mathrm{ST}}$ where $U_{\mathrm{ST}} = [0,L-1]^2\times [0,L]^2\times [0,T-1]$ and $B_{\mathrm{ST}}$ is  the union of four hyperplanes defined by the restriction $v_3\in \{0,L\}$ and  $v_4\in \{0,L\}$.

Errors form a surface in (4+1)-dimensional space-time. Let $E_e(t) = 1$ if the stabilizer $S^X_e$ is measured wrongly during measurement round $t\in\{0,1,\dots,T-1\}$ (where $T$ labels the total number of measurement rounds) and zero otherwise. We assume that the last round of measurements is perfect, $E_e(T-1)=0$.  Let $E_f(t)=1$ if the qubit corresponding to the face $f$ undergoes an error between measurement rounds $t-1$ and $t$ and zero otherwise. The error surface is given by $\tilde f_{\text{error}} = \sum_{\tilde{f}} E_{\tilde{f}} \tilde{f}$ where the coefficients $E_{\tilde{f}}$ are either related to qubit errors, $E_{\tilde f_{\{i,j\}}(\vec v +t\vec a_5)}= E_{f_{\{i,j\}}(\vec v)}(t)$ (for $i\leq4$ and $j\leq 4$), or measurement errors, $E_{\tilde f_{\{i,5\}}(\vec v +t\vec a_5 )}= E_{e_{\{i\}}(\vec v)}(t)$, depending on the orientation of $\tilde f$. 

If $\tau_e(t)\in \{0,1\}$ denotes the outcome of the faulty measurement of the stabilizer $S_e^X$ at round $t$, one has $\tau_e(t) = \sigma_e(t) + E_e(t)$. The syndrome curve $\tilde e_{\text{synd}} =  \sum_{\tilde{e}} \sigma_{\tilde{e}} \tilde{e}$ is a formal sum of edges in (4+1)-dimensional space-time with coefficients $\sigma_{\tilde{e}}$ given by:
\begin{align} \nonumber
 \sigma_{\tilde{e}_{\{i\}}(\vec v +t\vec a_5 )} &:= \\ \label{eq:m1}
 \tau_{e_{\{i\}}(\vec v)}&(t) - \tau_{e_{\{i\}}(\vec v)}(t-1) \ \ \text{ for } \ \ i<5\ \ ,\\ \label{eq:m2}
 \sigma_{\tilde{e}_{\{5\}}(\vec v +t\vec a_5)} &:= \sum_{e:\vec v \in e} \tau_e(t) = \sum_{e:\vec v \in e} E_e(t)  \ \ .
\end{align}
Intuitively, the first equation lets syndrome be non-zero when the regular syndrome which detects qubit errors changes from step $t-1$ to $t$. This change can occur either due to a qubit error or due to a  measurement error. The second equality in Eq.~\eqref{eq:m2} follows from the linear dependency of stabilizers, given in Eq.~\eqref{eq:sx_lin_dep}, and thus this non-zero syndrome heralds a measurement error. 

Note that by construction of the space-time cellular complex, the boundaries corresponding to the time directions are smooth. This is not the case if the last measurement round is faulty. The errors $E_e(T-1)$ can give rise to non-trivial syndrome on edges of the form $\tilde{e}_{\{5\}}(\vec v +(T-1)\vec a_5)$ which are not contained in $U_{\text{ST}}$. Including these edges gives rise to a rough $v_5=T$ boundary which can formalized by the cellular complex $U_{\mathrm{rough\;ST}} = [0,L-1]^2\times [0,L]^2\times [0,T]$ and $B_{\mathrm{rough\;ST}}\subset U_{\mathrm{rough\;ST}}$, being the union of five hyperplanes defined by the restriction $v_3\in \{0,L\}$, $v_4\in \{0,L\}$ and $v_5=T$.
\begin{figure}[hbt]
\centering
 \includegraphics[width = 0.45\textwidth]{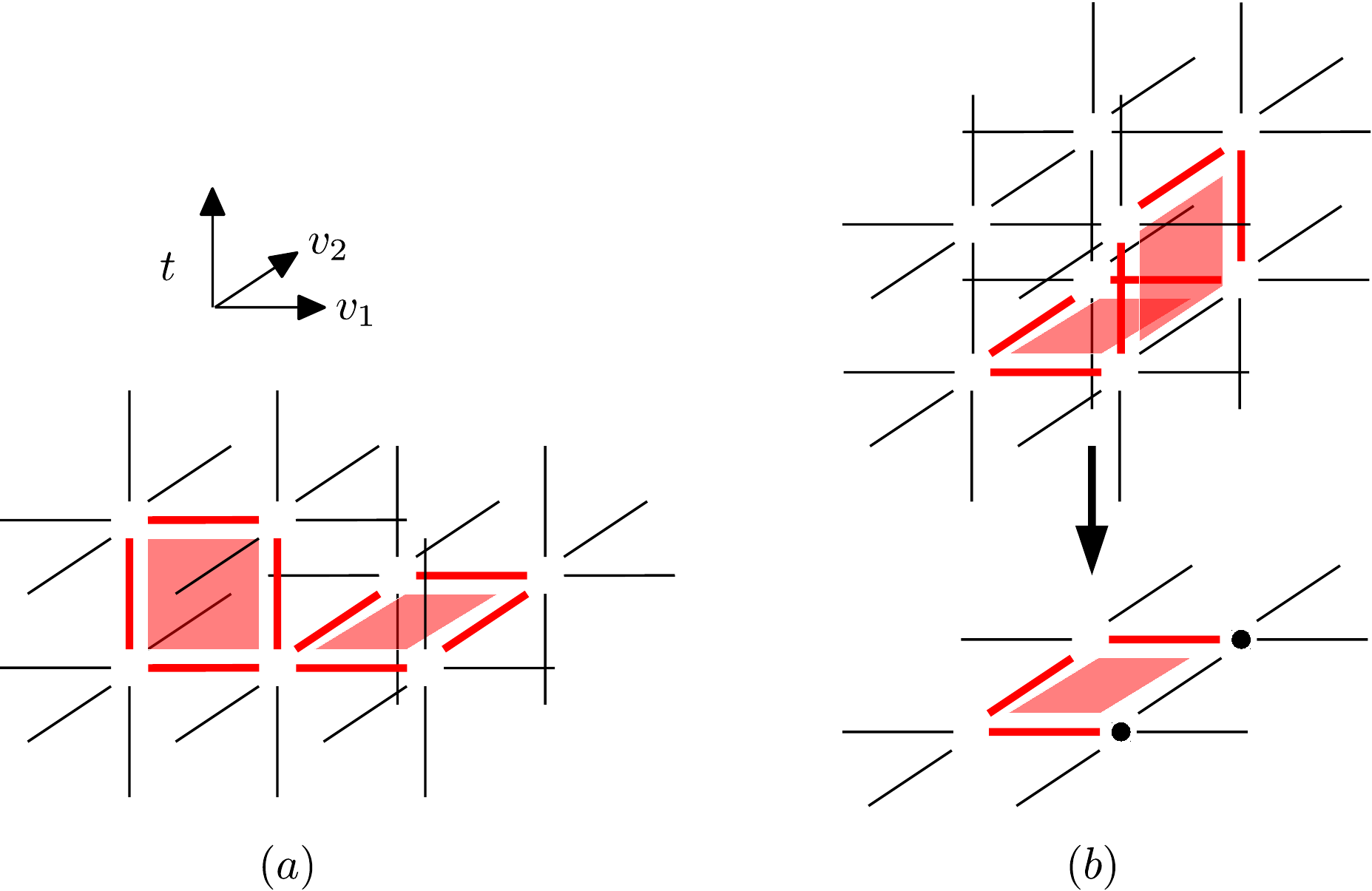}
 \caption{\label{fig:boundaries}(color online) Portion of the cross-section of the (4+1)-dimensional hypercube, containing two spatial directions and one (vertical) time direction. Panel (a): isolated qubit and measurement errors are depicted by red squares. Their boundary, corresponding the red lines, is the space-time syndrome curve $\tilde{e}_{\text{synd}}$. Panel (b): overlapping qubit and measurement error (upper figure) giving rise to an open syndrome curve ${e}_{\text{synd}}$ (lower figure).}
\end{figure}

It remains to check that the boundary of $\tilde f_{\text{error}}$ is indeed $\tilde e_{ \text{synd}}$. We argue that this is the case for single qubit errors or single measurement errors, see Fig.~\ref{fig:boundaries}. Then, by linearity, $\partial_2(\tilde f_{\text{error}}) = \tilde e_{\text{synd}}$ will hold for all error surfaces.
Assume there is an error on the qubit associated with face $f_{\{i,j\}}(\vec v )$ in the time interval $(t-1,t)$, i.e. $\tilde f_{\text{error}} = \tilde f_{\{i,j\}}(\vec v +t\vec a_5)$. The corresponding syndrome curve has coefficients $\sigma_{\tilde{e}}=0$ for any edge $\tilde{e}$ which is either oriented in the time direction (since there are no measurement errors) or which is not contained in the hyperplane $v_5 = t$ due to Eq.~\eqref{eq:m1}. The remaining coefficients satisfy $\sigma_{\tilde{e}} = \tau_{e}(t)$ which is non-zero if and only if $e\subset f_{\{i,j\}}(\vec v )$. It can be straightforwardly checked that this is exactly the boundary of $\tilde f_{\text{error}}$. Alternatively, assume there is a measurement error of the stabilizer associated with edge $e_{\{i\}}(\vec v)$ at time $t$, i.e. $\tilde f_{\text{error}} = \tilde f_{\{i,5\}}(\vec v +t\vec a_5 )$. Since now  $\tau_{{e}}(t) = E_e(t)$ (no qubit errors) we have, due to Eq.~\eqref{eq:m1}, that  $\sigma_{\tilde{e}_ {\{i\}}(\vec v +t\vec a_5)} =  \sigma_{\tilde{e}_{\{i\}}(\vec v +(t+1)\vec a_5)} = 1$. And since $\vec v \in e_{\{i\}}(\vec v)$ and $\vec v +\vec a _i\in e_{\{i\}}(\vec v)$ we have, due to Eq.~\eqref{eq:m2}, that $\sigma_{\tilde{e}_{\{5\}}(\vec v +t\vec a_5)} =  \sigma_{\tilde{e}_{\{5\}}(\vec v +\vec a _i+t\vec a_5)} = 1$. These coefficients exactly correspond to those edges contained in $\tilde f_{\{i,5\}}(\vec v +t\vec a_5 )$. It follows that the faulty-measurement minimum-weight decoding problem for the tesseract code is the problem of finding a minimal surface given its boundary in (4+1)-dimensional space-time. In Appendix~\ref{A:HD} we formulate this mapping quite generally for surface codes in $D$ dimensions.

\subsection{Gate-based error model}
\label{sec:gb}
In order to fairly compare the performance of the tesseract code with the surface code, we also consider a gate-based error model \cite{FSG:gatebased,RH:gatebased}, as opposed to a phenomenological error model. Every round of measurements consists of (1) ancilla preparation, (2) eight rounds of CNOT gates applied in parallel and (3) ancilla measurements. After $T-1$ rounds, a single round of non-faulty measurements is performed, without adding additional errors on the qubits. 

Ancilla qubits are defined for each stabilizer, i.e. there is an ancilla on each edge $e$ and on each cube $c$. Preparation is modeled by a perfect creation of the $\ket{+}$ state (for $X$-stabilizers) or the $\ket{0}$ state (for  $Z$-stabilizers), followed by a phase flip or bit flip channel with probability $p$. Ancilla measurement is modeled by a perfect measurement in the $X$ basis (for $X$-stabilizers) or the $Z$ basis (for $Z$-stabilizers), followed by a classical bit flip channel on the obtained measurement data, with probability $p$. 
During both preparation and measurement, data qubits undergo depolarizing errors with probability $p$: $\rho \mapsto (1-p) \rho + \frac{p}{3}(X\rho X + Y\rho Y + Z\rho Z)$. 

The CNOTs for the parity check circuits for the $X$- and $Z$-stabilizers, can, similar as for the toric code, see e.g. \cite{fowler+:review}, be fully run in parallel. This requires 8 rounds of CNOTs which is the minimal number of CNOT rounds to collect the entire syndrome since the qubit degree of the code, i.e. the number of parity checks that a qubit participates in, is 8. CNOTs are performed between data and ancilla qubits. Ancillas corresponding to edges are always the control qubit ($X$-stabilizer) whereas ancillas corresponding to cubes are always the target qubit ($Z$-stabilizer). Ancilla or data qubits, on which no CNOT acts during a round, undergo depolarizing error with probability $p$. The CNOT gates are modeled by a perfect CNOT followed by the channel $\rho \mapsto (1-p) \rho + p/15\sum_{i=1,j=1\colon ij\neq 11}^{4,4} P_i^1 P_j^2\rho P_i^1 P_j^2$ where $\rho$ is the density matrix of the two qubits on which the CNOT acts and $P_1 = \mathbb{I}$.  

To explain which CNOT is performed in which round we associate with each such gate a direction, being the direction of the location of the ancilla qubit (edge or cube) with respect to the data qubit (face) on which the CNOT acts. There are 8 such directions. During a single round all CNOT gates oriented along a certain direction are performed. This ensures that there are never multiple CNOTs acting on the same qubits. The order of these 8 directions indicates for each qubit on a face the order in which the qubit interacts with the ancillas which are on the 4 edges surrounding the face and the 4 cubes which contain the face. In more detail, let the 8 directions be given by $(-1)^n\vec a_k$ specified by $n\in\{0,1\}$, $k\in\{1,2,3,4\}$. During a single round, labeled by direction $(-1)^n\vec a_k$, CNOT gates are applied between data qubits corresponding to faces $f_{\{i,j\}}(\vec v)$ (wlog, let $i\neq k$) and ancilla qubits corresponding to either cubes $c_{\{i,j,k\}}(\vec v +n\vec a_k)$ if  $j\neq k$ or edges $e_{\{i\}}(\vec v +(1-n)\vec a_k)$ if $j=k$, and if the corresponding $c$ and $e$ are elements of $C_L$ and $E_L$, respectively. The ordering of the different directions is given by  $[-\vec a_1 , -\vec a_2 ,-\vec a_3 ,-\vec a_4 ,\vec a_4 ,\vec a_3 ,\vec a_2 ,\vec a_1 ]$. In Appendix~\ref{A:cnot} we verify that with this schedule the execution of $X$- and $Z$-stabilizer measurements is not interfering. Note that due to this ordering the error model is not invariant under interchanging the primal and dual lattices. Hence the error rate for logical $\overline X$ and logical $\overline Z$ errors, and so the thresholds, could be different (we will only consider Z errors in Section \ref{sec:RG}).

\section{Decoding the tesseract code}
\label{sec:decoder}
As described above, minimum-weight decoding is equivalent to finding a minimal surface corresponding to a given curve in a five-dimensional space-time. Although this strictly only holds for a phenomenological error model, we will also use this strategy to correct for errors induced by the gate-based error model. One can generally ask about the complexity of the problem of finding a minimal (facial) surface in a $D$-dimensional hypercubic lattice, with $D\geq 3$, given its one-dimensional boundary. Results in \cite{thesis:sullivan} suggest that there is an efficient algorithm in three dimensions but one does not expect this to generalize to four or five dimensions. It can be noted that the minimal surface with a given boundary does not need to be orientable, i.e. one can have errors on faces which form a M\"obius strip. We address the complexity problem by introducing an efficient decoding scheme based on renormalization: this scheme is not guaranteed to find the minimal surface but our results demonstrate that it performs sufficiently well for the application of quantum error correction.

Before giving the decoder based on renormalization, we describe a single-shot decoder which repairs the faulty measurement data before attempting to find a minimal surface in 4D.

\begin{figure}[bt]
\centering
 \includegraphics[width = 0.3\textwidth]{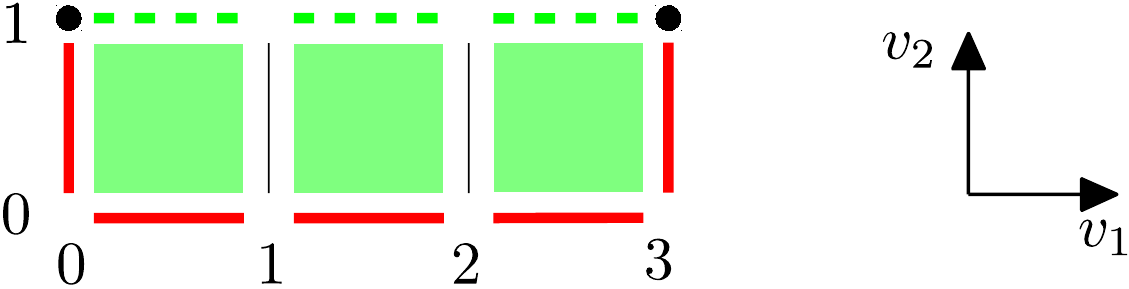}
 \caption{\label{fig:min}(color online) Illustration of a non-optimal two-step single-shot decoding process. The five red lines indicate a measured syndrome curve $e_{\text{synd}}$. The solution $f_{\text{cor}} = 0$, $e_{\text{cor}} = e_{\text{synd}}$ to Eq. \eqref{eq:global} has total Hamming weight 5. First pairing the end points $\partial_1(e_{\text{synd}})$ (black dots) results in $e_{\text{cor}}$ being the three dotted green lines. The corresponding $f_{\text{cor}}$ are the three green faces. This solution has $|f_{\text{cor}}|+|e_{\text{cor}}|=6$, which is larger than the optimal solution. }
\end{figure}

\subsection{Single-shot Repair-Syndrome Decoder}
\label{sec:ss}

Without any measurement errors, the threshold of the four-dimensional toric code is upper-bounded by $11.003\%$ \cite{takeda}. The question is whether the process of accurately correcting erroneous syndrome data can have a threshold of the same order of magnitude. This would make the whole decoding process for faulty measurements for the tesseract code have a threshold which is substantially larger than the less than $3\%$ of the surface code. It has been shown in \cite{bombin:selfcorrection} that such single-shot decoder which repairs the syndrome has an actual threshold.

When the syndrome measurements are faulty, it will not be the boundary of some surface in four dimensions. Let $e_{\text{error}}$ be the edge set corresponding to all measurement errors $e_{\text{error}} = \sum_eE_ee$. The erroneous syndrome curve at some fixed time $t$ can be written as:
\begin{align}\label{eq:global}
 e_{\text{synd}} = \sum_e \tilde \tau_e e = \partial_2(f_{\text{error}}) + e_{\text{error}}\ \ .
\end{align}
This should not be confused with $\tilde e_{\text{synd}}$ which is the boundary of a surface in (4+1)-dimensional space-time, see Fig.~\ref{fig:boundaries}(b) which depicts the relation between the two.
Given the measurement data, a single-shot repair-syndrome decoder aims to find the most likely correction $f_{\text{cor}}$ and $e_{\text{cor}}$ such that $e_{\text{synd}} =  \partial_2(f_{\text{cor}}) + e_{\text{cor}}$.
Consider the strategy of the following decoder which consists of two steps. In a first step the syndrome is ``repaired''. Due to $\partial_1\circ\partial_2 =0$ we have that $\partial_1 (e_{\text{synd}}) = \partial_1(e_{\text{cor}})$. The correction $e_{\text{cor}}$ is a set of edges having the same endpoints as $e_{\text{synd}}$. Moreover, the decoder will search for a correction which minimizes $|e_{\text{cor}}|$. This translates to a matching problem, matching the endpoints of $e_{\text{synd}}$, and can be done by Edmonds' efficient minimal-weight perfect matching algorithm. The corresponding repaired curve $e_{\text{synd}} + e_{\text{cor}}$ is closed. Since $\dim(H_1) = 0$ (no nontrivial closed curves) for the tesseract cellular complex, all closed curves are the boundary of some surface and hence $e_{\text{synd}} + e_{\text{cor}}$ can be used in a second step to find such a surface. This can for example be done using the renormalization group decoder, see Section~\ref{sec:decode}.

\begin{figure}[bt]
\centering
 \includegraphics[width = 0.35\textwidth]{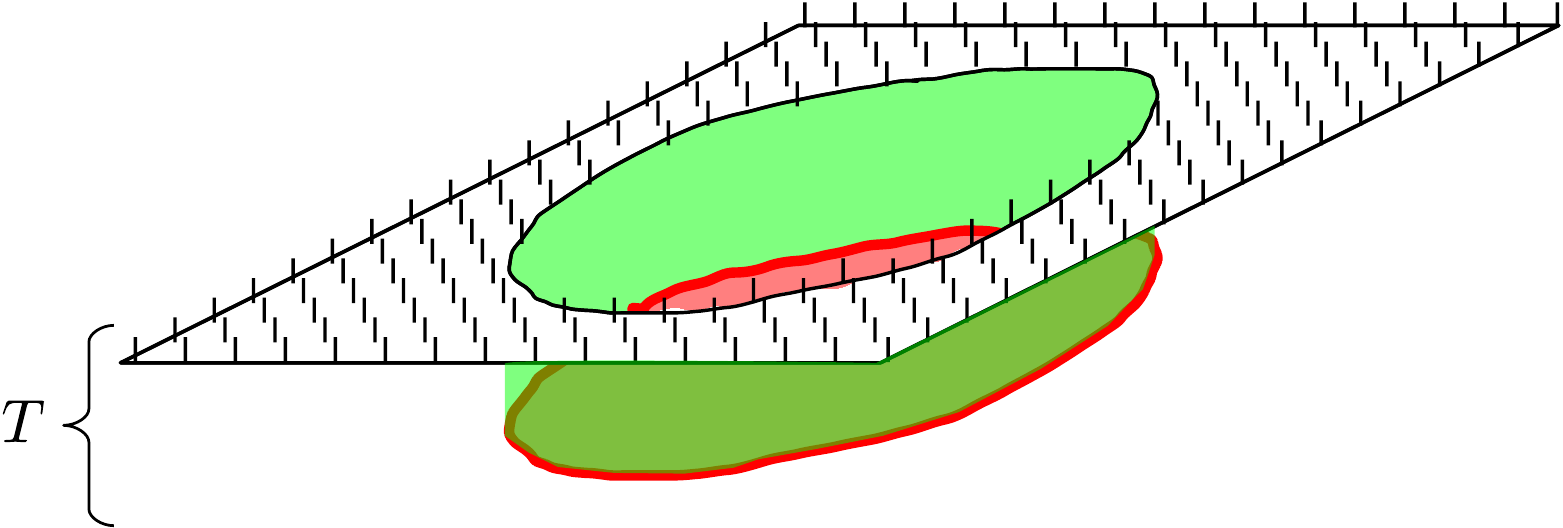}
 \caption{\label{fig:single-step}(color online) Illustration of how a decoder is unable to remove a patch of qubit errors (red). The syndrome (red curve) is also the boundary of the green rim which corresponds to $T$ rounds of measurement errors. Note how the green rim terminates on a rough time boundary and hence \textit{only} has the syndrome as its boundary. If the red patch is sufficiently large and $T=O(1)$ then the minimal surface will always be the green rim since the boundary of the patch grows slower than its volume. In other words the error is interpreted as measurement error.}
\end{figure}

Note that this strategy of first minimizing $|e_{\text{cor}}|$ and then separately minimizing $|f_{\text{cor}}|$ can give a suboptimal result as compared to minimizing both quantities simultaneously. Fig.~\ref{fig:min} gives an example illustrating the issue. In addition, one can observe that although the distance of the of the tesseract code scales with $L^2$, specific errors of Hamming weight $O(L)$ can lead to logical failure. Consider a curve of length $O(L)$, dividing a surface corresponding to a logical operator in half. If such a curve corresponds to erroneous measurements $e_{\text{error}}$, its minimal surface is almost half of a logical operator. Hence, if additionally $O(1)$ qubit errors occur on the other half the memory is corrupted. Based on this argument one in fact expects that the threshold of the single-shot repair-syndrome decoder for the tesseract code is upper bounded by the threshold of the line-logical operator of the $(d_1=3,d_2=1)$-surface code, see an elaboration of this heuristic argument in Appendix \ref{A:HD}. These arguments thus indicate that the optimal single-shot decoding threshold will be different than the optimal space-time decoding threshold for a $(d_1,d_2)$-surface code. In Section \ref{sec:perf_ss} we numerically study single-shot decoding for the 4D tesseract code and finds that its performance is indeed below the performance of the RG decoder in 5D.

One can ask whether for the tesseract code single-shot decoding in space-time with an $O(1)$ time-direction would lead to a single-shot decoder with a noise threshold. Even though the tesseract code is self-correcting, one can argue that this is unlikely to work due to arguments about the scaling of volumes vs. boundaries of volumes. One can imagine a sliding time-window as in \cite{DKLP} in which syndromes are processed within a window of size $T=O(1)$. As illustrated in Fig.~\ref{fig:single-step}, whenever $|f_{\text{error}}| > T|\partial_2( f_{\text{error}})|$, the area of the vertical surface connecting the syndrome to the future time boundary is smaller than that of the horizontal surface enclosed by the syndrome. Note that since the last measurement is faulty, the future time boundary is rough, as described in Section~\ref{sec:phen}, making it possible for a surface to terminate at this boundary. This means a bubble of qubit errors growing as a function of $L$ would be interpreted as repeated measurement error under minimum-weight decoding in this $O(1)$ window. Sliding the window forward by fixing the error found in the latter half of the time-window then simply carries the problem of the uncorrected qubit bubble forward to the next decoding round.


\subsection{Renormalization group decoder}
\label{sec:decode}

The renormalization group decoder aims to find a correction of minimal Hamming weight satisfying $\partial_2(f_{\text{cor }})= e_{\text{synd}}$ where $e_{\text{synd}}$ is a set of closed curves. The decoder works for any generalized surface code defined in Appendix~\ref{A:HD} having a surface-like logical $\overline{Z}$, i.e. $d_2 = 2$. The application here will be decoding errors on a hypercubic lattice obtained from $U,B$ or $U_{\text{ST}}$,$B_{\text{ST}}$ in four respectively five dimensions. Due the RG structure of the decoder, we will only describe it for hypercubic lattices of size $L(N)=2^N+1$ for some integer $N$, but the ideas could also be applied to lattices of different sizes.

\begin{figure}[tb]
\centering
 \includegraphics[width = 0.45\textwidth]{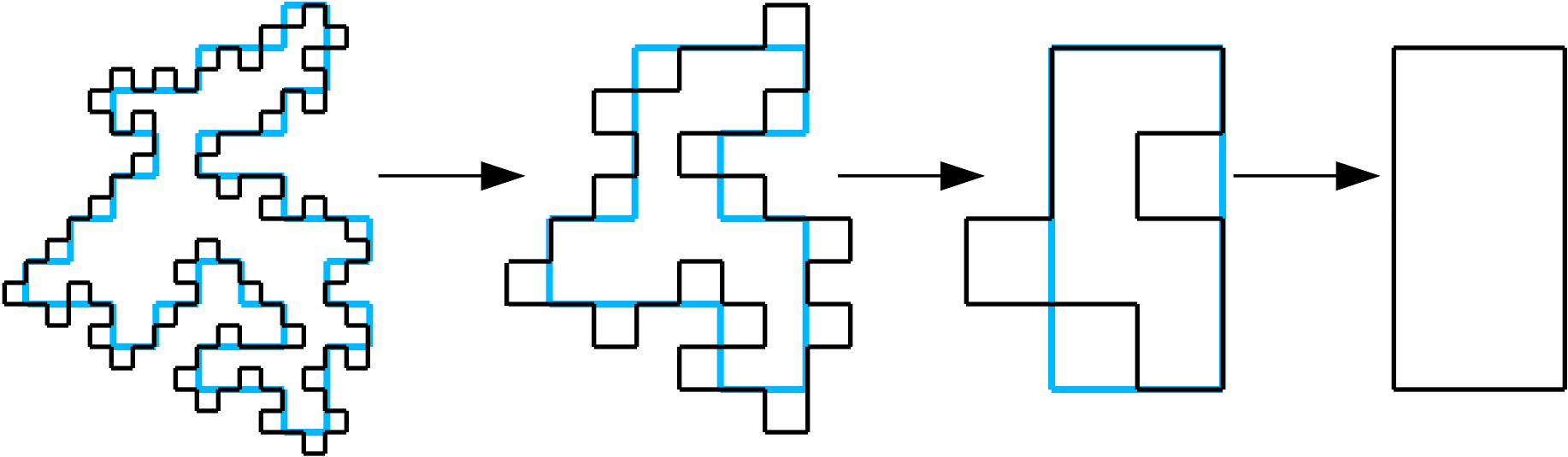}
 \caption{\label{fig:TODO}(color online) Illustration of the philosophy of the renormalization group decoder. By straightening out the syndrome curve one is able to find a corresponding surface on a larger scale. }
\end{figure}

The idea of the decoder is to straighten out the syndrome in a series of coarse-graining steps, see Fig.~\ref{fig:TODO}. In each step a partial correction $f_{\text{cg}}$ reduces the syndrome to a syndrome which is only supported on a smaller coarse-grained sublattice. The problem on the coarse-grained lattice can be identified with the same original decoding problem but now on a lattice with $L(N-1)$ and hence one can apply the same method to reduce the syndrome again. The coarse-grained sub-lattices can be best visualized for a low-dimensional lattice, see Fig.~\ref{fig:lattice}. In a last step, when the lattice can no longer be further coarse-grained, the decoding problem is solved by solving an integer programming problem \cite{gurobi}. In Section~\ref{sec:rg_cg} we define the sub-lattice. In Section~\ref{sec:rg_cor} we explain how we find the partial correction $f_{\text{cg}}$. The Matlab code for this algorithm can be found on GitHub \url{https://github.com/kduivenvoorden/tesseract_code}.

\begin{figure*}[!t]
\normalsize
\begin{align}\label{eq:mape}
 \Gamma_N^E (e_{\{i\}}(\vec v))=   & \left\{
 \begin{array}{ll}
  e_{\{i\}}(2\vec v) &\text{if }  i\in\{3,4\} \text{ and } v_i = 2^{N-1} \\
 e_{\{i\}}(2\vec v) + e_{\{i\}}(2\vec v+\vec a_i) &\text{else}
 \end{array}\right. \ \ ,\\ \label{eq:mapf}
 \Gamma_N^F (f_{\{i,j\}}(\vec v))= & \left\{
 \begin{array}{ll}
 f_{\{i,j\}}(2\vec v)  &\text{if }  i,j\in\{3,4\} \text{ and } v_i = v_j = 2^{N-1} \\
 f_{\{i,j\}}(2\vec v) + f_{\{i,j\}}(2\vec v+\vec a_i) &\text{else if }  j\in\{3,4\} \text{ and } v_j = 2^{N-1} \\
 f_{\{i,j\}}(2\vec v) + f_{\{i,j\}}(2\vec v+\vec a_j) &\text{else if }  i\in\{3,4\} \text{ and } v_i = 2^{N-1} \\
 f_{\{i,j\}}(2\vec v) + f_{\{i,j\}}(2\vec v+\vec a_i) 
 + f_{\{i,j\}}(2\vec v+\vec a_j) + f_{\{i,j\}}(2\vec v+\vec a_i+\vec a_j) &\text{else}
 \end{array}\right.\ \ .
\end{align}
\hrulefill
\vspace*{4pt}
\end{figure*}

\begin{figure}[t]
\centering
 \includegraphics[width = 0.45\textwidth]{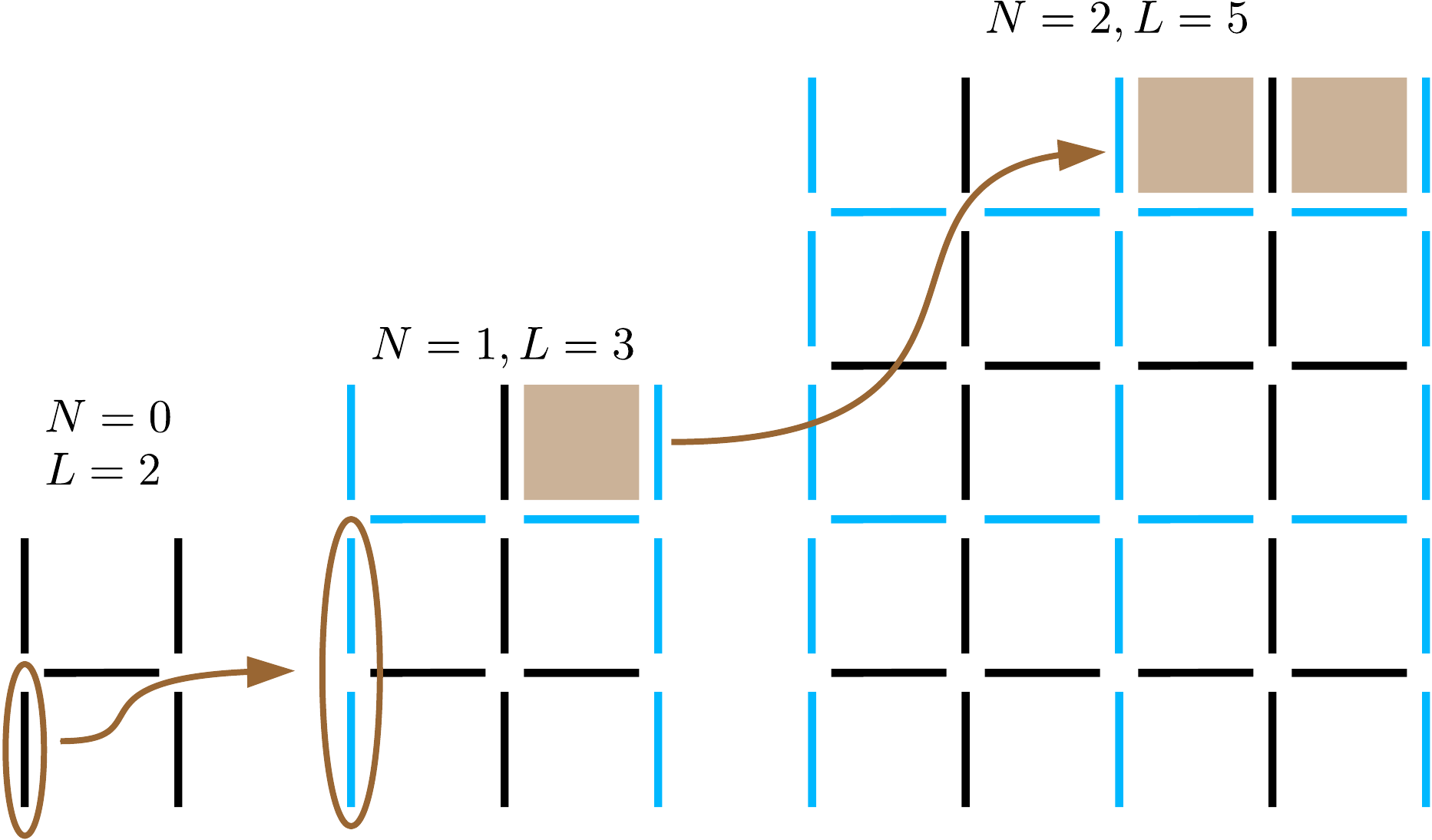}
 \caption{\label{fig:lattice}(color online) Edges of the surface code of length $L=2$ (left), $L=3$ (middle) and $L=5$ (right). The edges forming the coarse-grained lattice of the larger two codes, $L=3$ and $L=5$, are depicted in blue. These correspond to the image of $\Gamma^E$, i.e. the embedding of the smaller codes $L=2$ and $L=3$. An example of a mapping of an edge $\Gamma^E(e)$ and a face $\Gamma^F(f)$ is depicted in brown.}
\end{figure}

\subsubsection{Coarse-graining}
\label{sec:rg_cg}
The coarse-grained sublattice $E_L^{\text{cg}}$ is a subset of the edge set $E_L$ and contains edges which are incident on a vertex with even coordinates. These edges are either of the form $e_{\{i\}}(2\vec v)$ or of the form $e_{\{i\}}(2\vec v+\vec a_i)$. 
For lattice sizes $L(N) = 2^N+1$ we will denote the space of formal sums of edges in $E_L$ as ${C}_1(N)$ and the space of formal sums of edges in $E_L^{\text{cg}}$ as ${C}^{\text{cg}}_1(N)$. The latter can also be understood as the image of the coarse-graining map $\Gamma^E_N$ where $\Gamma^E_N \colon C_1(N-1) \rightarrow  C^{\text{cg}}_1(N)$ maps edges of a smaller tesseract code of size $L(N-1)$, into a larger code, of size $L(N)$. 
Similarly, faces in the smaller lattice are mapped to four faces of the larger lattice by the map $\Gamma^F_N\colon C_2(N-1) \rightarrow C_2(N)$. Concretely, for a hypercubic lattice with two rough boundaries, the action of $\Gamma_{N}^E$ and $\Gamma_{N}^F$ are given by Eqs.~\eqref{eq:mape} and \eqref{eq:mapf}. The basic statement (last line in both equations) is for the bulk of the lattice while the conditional statements ensure that some of the edges/faces at the rough boundary are mapped to only those edges and faces which are actually part of the lattice of size $L(N)$. From here onwards we will drop the subscript $N$ of $\Gamma_{N}^E$ and $\Gamma_{N}^F$. Note that $\Gamma^E$ is not surjective  since ${C}^{\text{cg}}_1(N)$ contains the edge $e_{\{i\}}(2\vec v)$ and the edge $e_{\{i\}}(2\vec v+\vec a_i)$, while the image of $\Gamma^E$ is only spanned by sums of two such edges (with the exception of some rough-boundary edges). Nevertheless, any closed curve contained in ${C}^{\text{cg}}_1(N)$ is also contained in the image of $\Gamma^E$.

\begin{figure*}[bt]
\centering
 \begin{subfigure}[b]{.3\textwidth}
 \includegraphics[width = 1\textwidth]{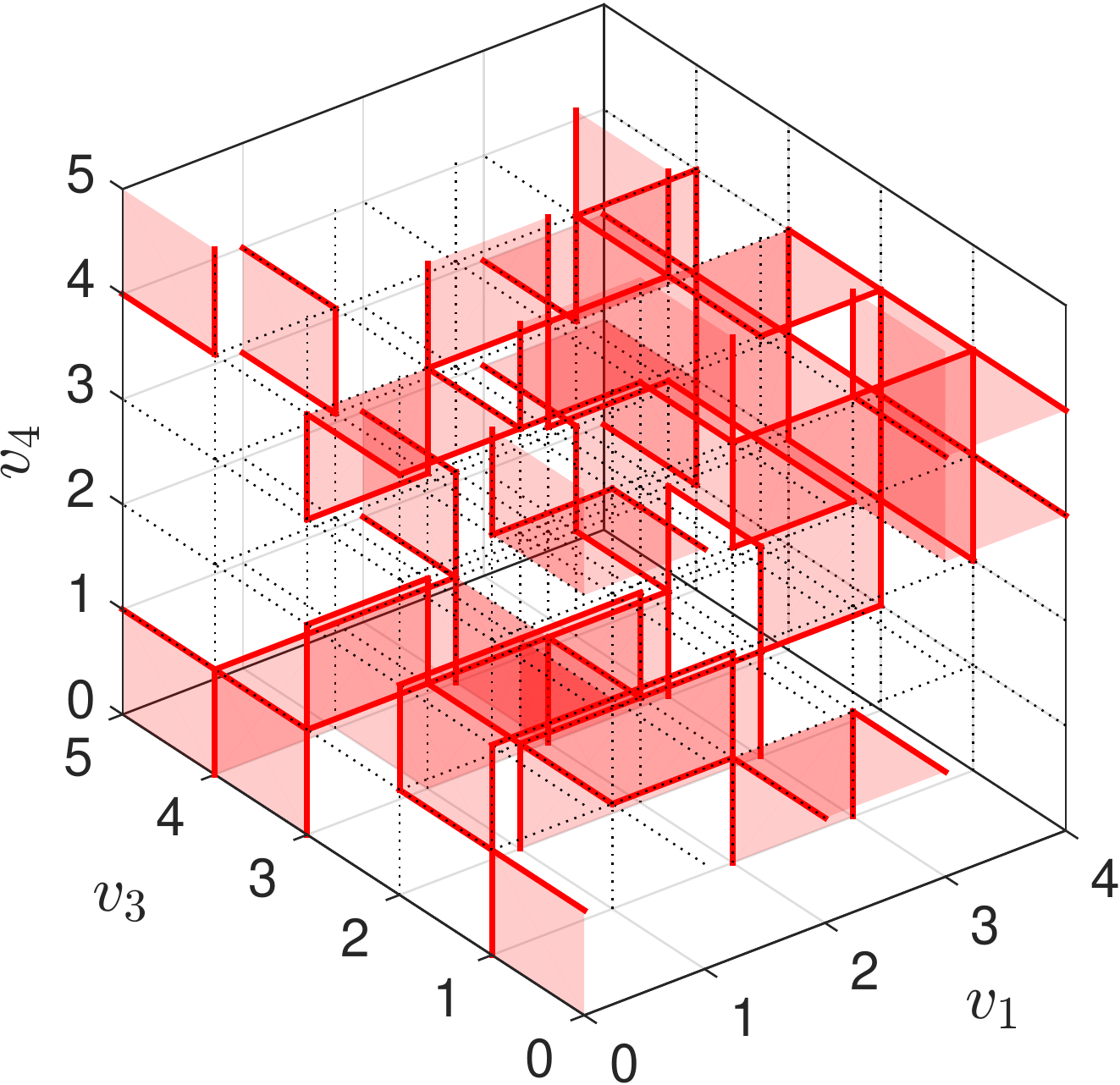}
 \caption{$L=5$ syndrome}
 \label{fig:4a}
 \end{subfigure}
 ~
 \begin{subfigure}[b]{.3\textwidth}
 \includegraphics[width = 1\textwidth]{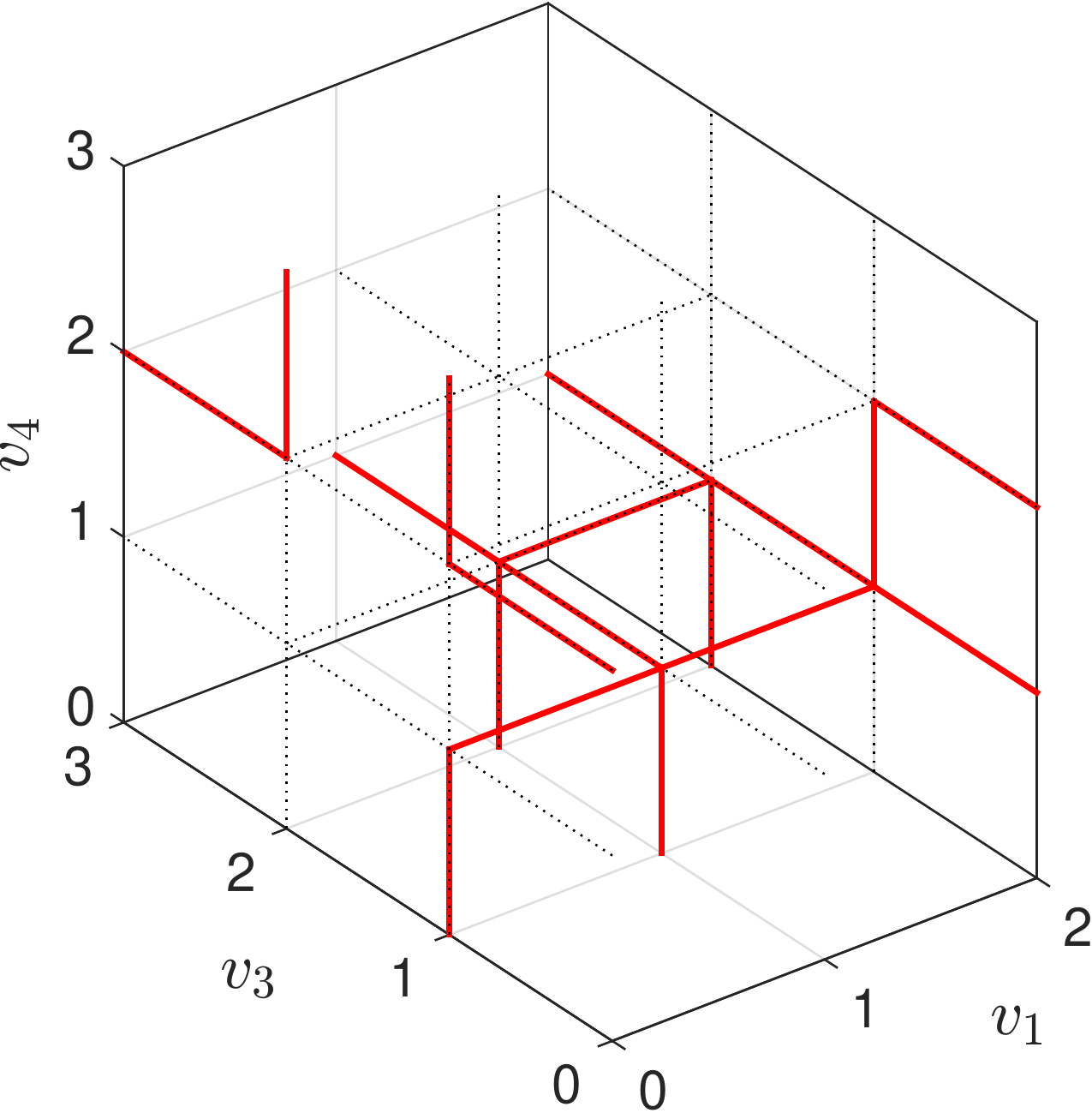}
 \caption{$L=3$ syndrome}
 \label{fig:4b}
 \end{subfigure}
 ~
 \begin{subfigure}[b]{.3\textwidth}
 \includegraphics[width = 1\textwidth]{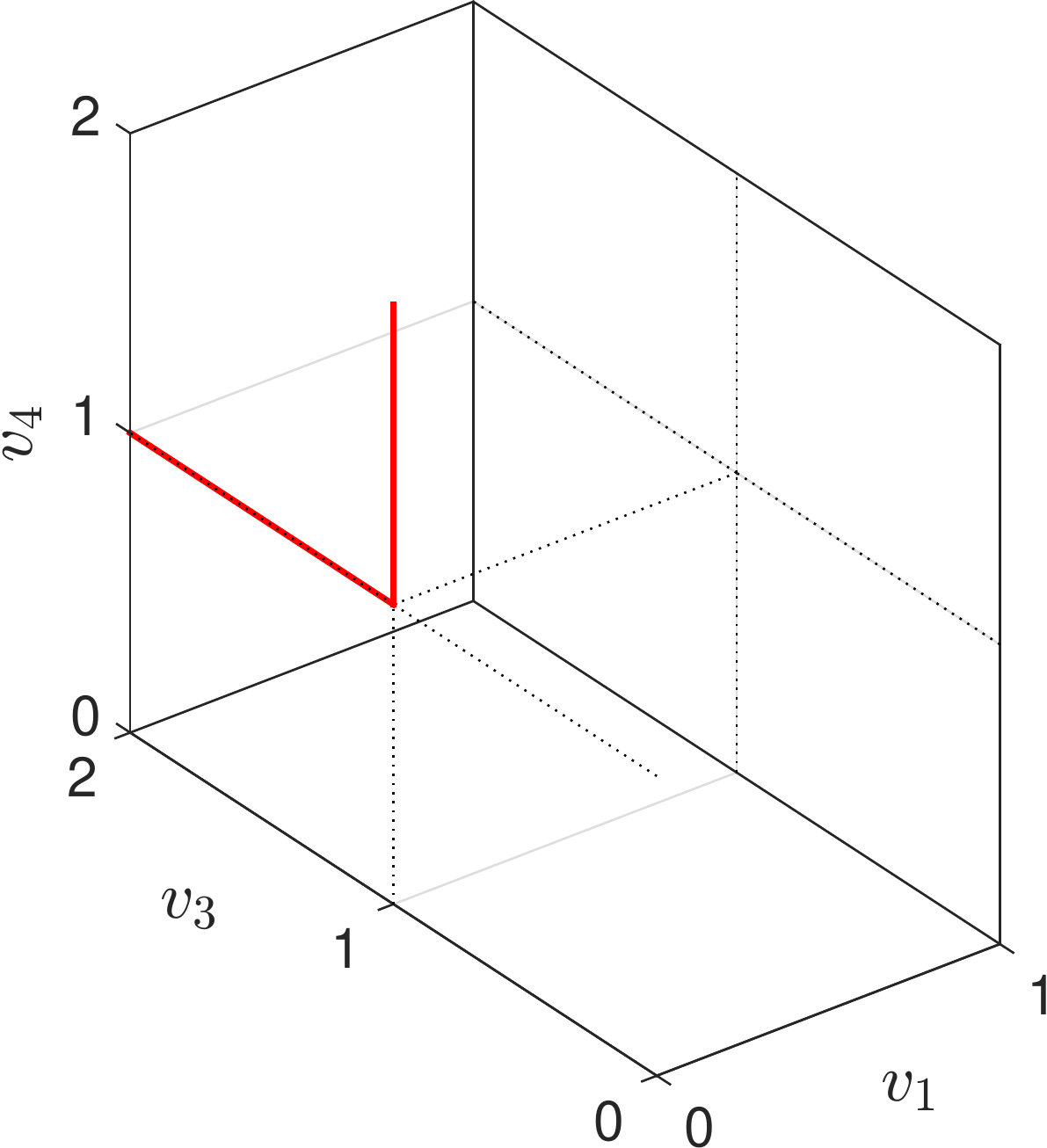}
 \caption{$L=2$ syndrome}
 \label{fig:4c}
 \end{subfigure}
 
 \begin{subfigure}[b]{.3\textwidth}
 \includegraphics[width = 1\textwidth]{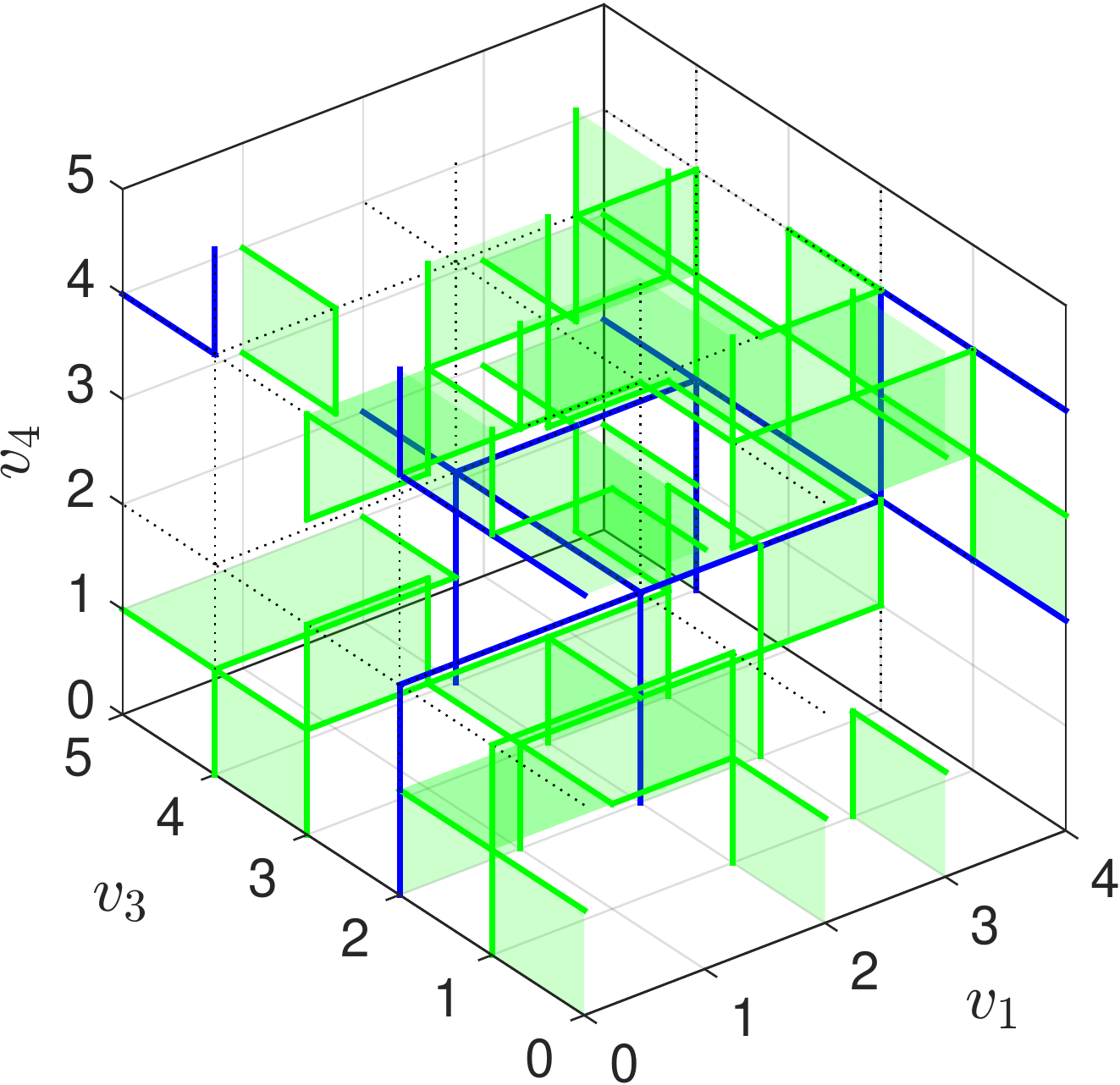}
 \caption{$L=5$ correction}
 \label{fig:4d}
 \end{subfigure}
 ~
  \begin{subfigure}[b]{.3\textwidth}
 \includegraphics[width = 1\textwidth]{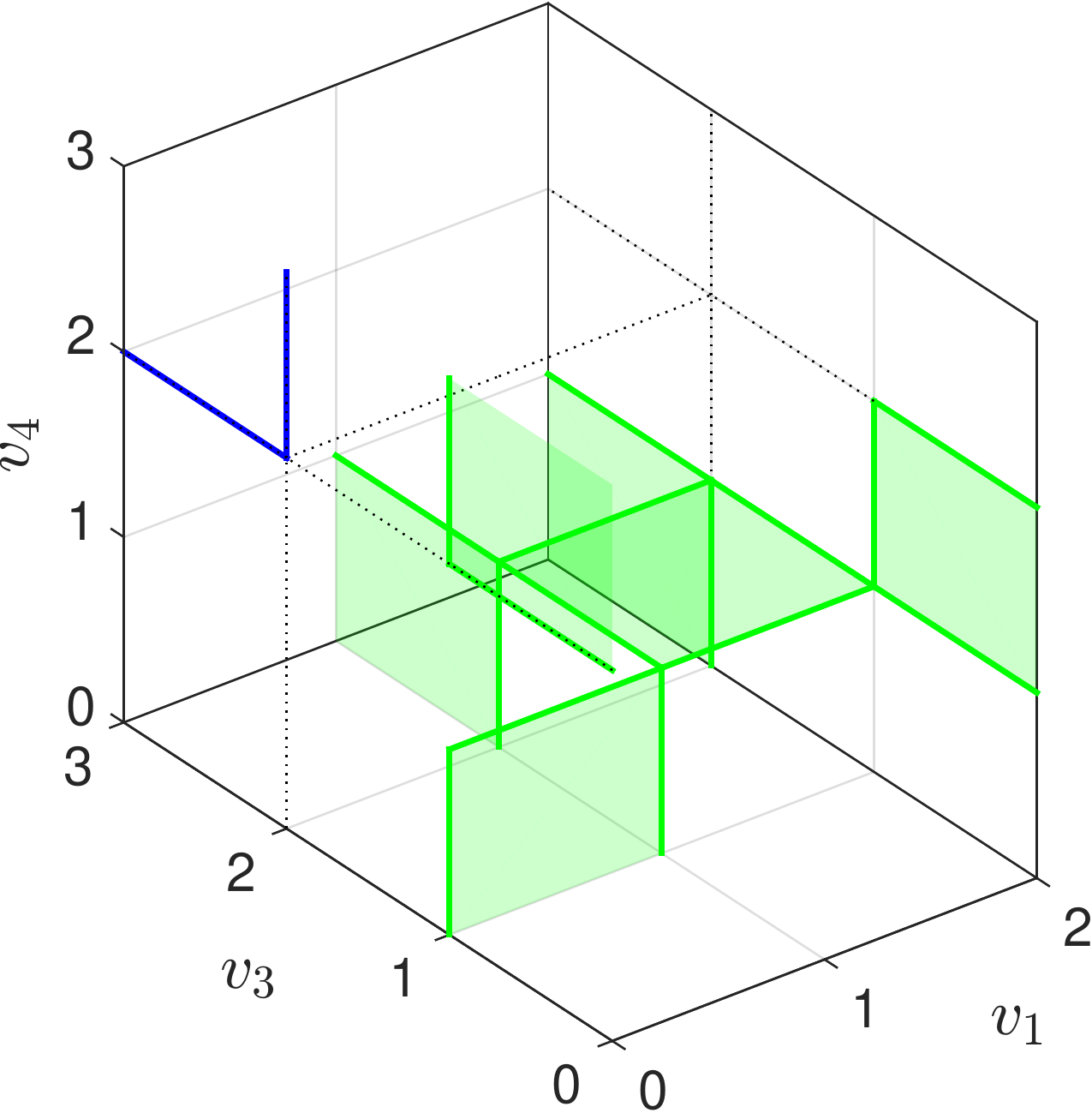}
 \caption{$L=3$ correction}
 \label{fig:4e}
 \end{subfigure}
 ~
  \begin{subfigure}[b]{.3\textwidth}
 \includegraphics[width = 1\textwidth]{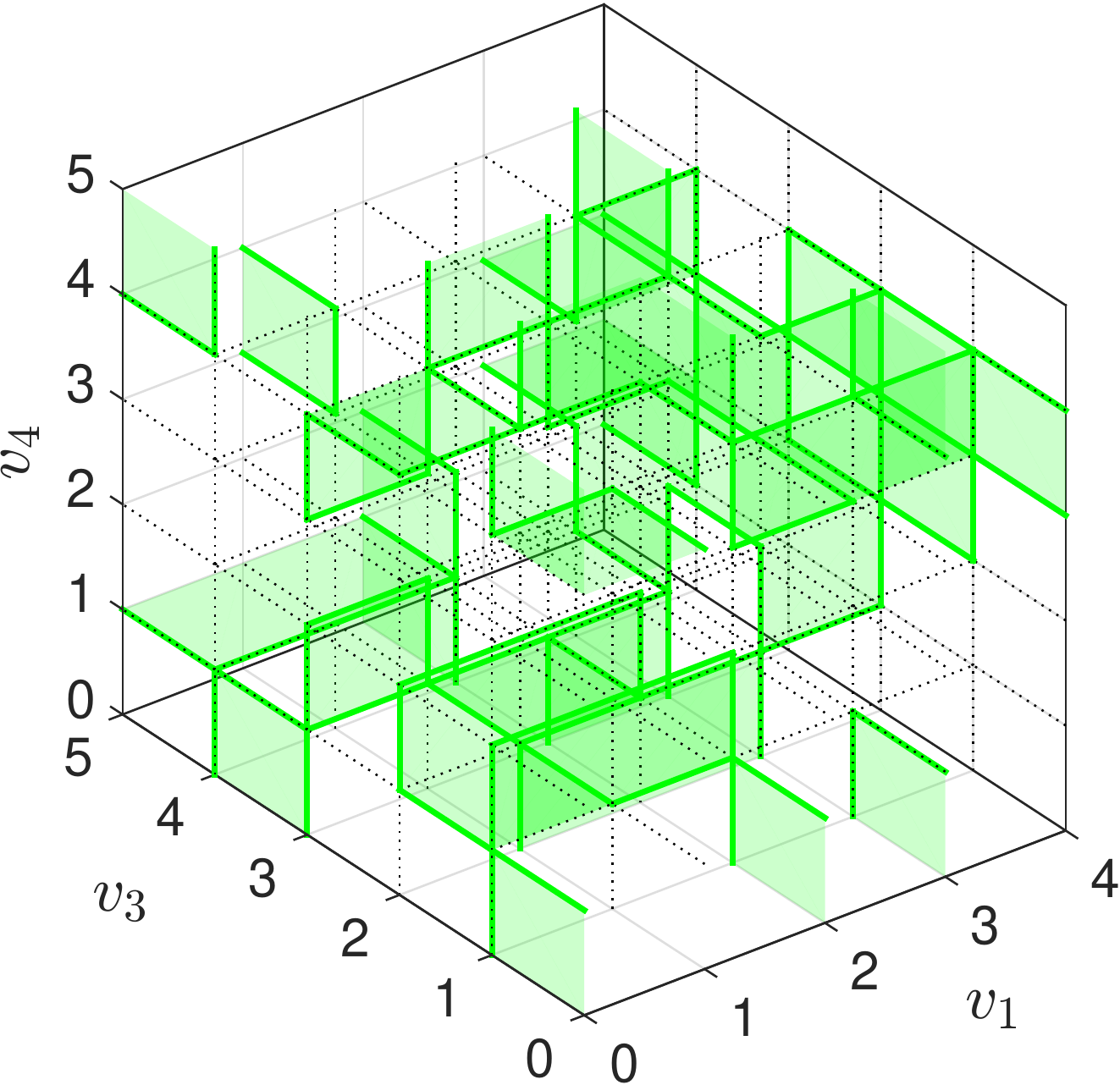}
 \caption{Result}
 \label{fig:4f}
 \end{subfigure}
 \caption{\label{fig4}(color online) The decoder in action: Panel a) indicates qubit errors on red faces and the corresponding syndrome $e_{\text{synd}}$ in red. Panel d) indicates a set of faces $f_{\text{cg}}$ in green. The difference $e_{\text{synd}}+\partial_2 (f_{\text{cg}})$ is depicted in blue, and corresponds to the remaining syndrome after applying this correction. The remaining syndrome can be mapped to an $L=3$ code and the corresponding panels b), e) and c) depict a second coarse-graining procedure. The remaining syndrome depicted in panel c) is corrected in a last step. Panel f) indicates all the faces corresponding to qubits which have been corrected during the full procedure.}
\end{figure*}

For the syndrome $e_{\text{synd}}$ on a lattice of size $L(N)$ we aim to find a partial correction $f_{\text{cg}} \in C_2(N)$ such that $e_{\text{synd}} + \partial_2 (f_{\text{cg}}) \in {C}^{\text{cg}}_1(N)$.  In words: we aim to reduce the syndrome to having only support on the coarse-grained lattice.
Since $e_{\text{synd}} + \partial_2 (f_{\text{cg}})$ is some set of closed curves, it can be written as $\Gamma^E(e_{\text{synd}}^{\text{red}})$, i.e. it can be identified to a reduced syndrome $e_{\text{synd}}^{\text{red}}$ on a smaller lattice, of size $L(N-1)$. 
When one solves the problem on the smaller lattice, that is, finds a $f_{\text{cor}}^{\text{red}}\in C_2(N-1)$ such that $\partial_2 (f_{\text{cor}}^{\text{red}}) = e_{\text{synd}}^{\text{red}}$, one can map it back to the original lattice, $f_{\text{cor}}= \Gamma^F({f}_{\text{cor}}^{\text{red}})$ . The total correction is hence $f_{\text{tot}} =  f_{\text{cor}} +  f_{\text{cg}}$ and obeys $\partial_2 (f_{\text{tot}})  = e_{\text{synd}}$ due to the commutation of the coarse-graining map $\Gamma$  with the boundary operator in the sense that
\begin{align} \nonumber
 \Gamma^E \circ \partial_2 =  \partial_2 \circ \Gamma^F\ \ .
\end{align}
The problem of finding the solution $f_{\text{cor}}^{\text{red}}$ on the smaller lattice can, by applying the same recursive step, be reduced to an even smaller lattice etc. Two coarse-graining steps, used to solve a $L=5$ cubic code, are depicted in Fig.~\ref{fig4}. 

Optimally, the decoder should find $f_{\text{cg}}$ and $f_{\text{cor}}^{\text{red}}$ such that Hamming weight of $f_{\text{tot}} = f_{\text{cg}}+{f}_{\text{cor}}$ is minimized.  Using the notation $a\cdot b := \sum _f \alpha_f \beta_f\in\mathbb{R}$ where $a = \sum_f \alpha_f f$ and  $b= \sum_f \beta_f f$, we can formally rewrite the Hamming weight as 
\begin{align} \nonumber
|f_{\text{tot}}|   = |f_{\text{cg}}| +  w\cdot{f_{\text{cor}}},
\end{align}
with a weight vector $w=\sum_f (-1)^{\alpha_f} f$ with $f_{\text{cg}}=\sum_f \alpha_f f$, $\alpha_f \in \{0,1\}$. Instead of minimizing $|f_{\text{tot}}|$, the decoder minimizes $|f_{\text{cg}}|$ in a coarse-graining step and then minimizes $w\cdot{f_{\text{cor}}}$ in subsequent steps. 
The quantity $w\cdot{f_{\text{cor}}}$ can be rewritten as $ {w} \cdot \Gamma^F(f_{\text{cor}}^{\text{red}}) =[ (\Gamma^F)^T  (w)] \cdot f_{\text{cor}}^{\text{red}}$. Note that this mapping of weights is simply due to the equivalence of flipping a qubit corresponding to a face on the coarse-grained lattice, to flipping the qubits corresponding to the related faces of the original lattice. Thus in the next step, the minimization problem is to find a $f_{\text{cor}}^{\text{red}}$ which obeys $\partial_2 (f_{\text{cor}}^{\text{red}}) = e_{\text{synd}}^{\text{red}}$ while minimizing 
$w^{\text{red}} \cdot f_{\text{cor}}^{\text{red}}= [(\Gamma^F)^T (w)]\cdot f_{\text{cor}}^{\text{red}}$.

\begin{figure}[bt]
\centering
 \includegraphics[width = 0.45\textwidth]{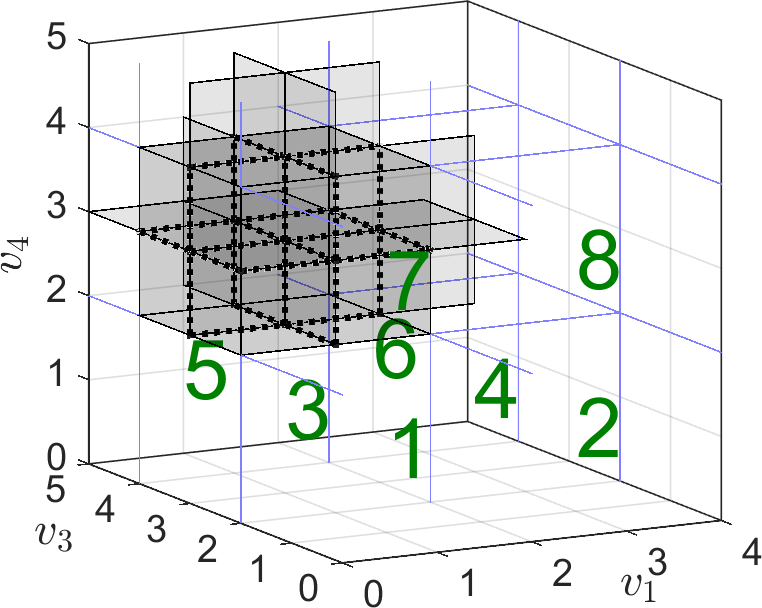}
 \caption{\label{fig:box}(color online) Illustration of the box $\mathcal{B}({\vec v })$, corresponding to vertex $\vec v = \vec a_1 + 3\vec a_3 +3\vec a_4$.  The black dotted lines are its edges, $E({ \vec v})$, and the gray squares its faces $F({\vec v })$. It is the 9-th box in terms of the ordering $\Omega(\vec v)$. The green numbers indicate the location and the order of boxes which are treated before this box by the algorithm.}
\end{figure}


\subsubsection{Correcting}
\label{sec:rg_cor}
In order to find a correction $f_{\text{cg}}$ such that $e_{\text{synd}} + \partial_2(f_{\text{cg}})$ is in the image of $\Gamma^E$, while minimizing $w\cdot f_{\text{cg}}$, we divide the lattice into boxes containing a number of edges and faces. We consecutively treat the decoding problem for each such box by solving an integer linear programming problem. 
Complications arise due to the fact that the boxes overlap. On the one hand, solutions for a certain box can alter the integer program of the still to-be-solved boxes, and on the other hand, these solutions should not corrupt the solution of already-solved boxes.

For every vertex $\vec v$ with only odd coordinates, we define a box $\mathcal{B}({\vec v }) = \{E({\vec v }), F({\vec v })\}$, consisting of a subset $E({\vec v })\subset E_L$ of edges and a subset $F({\vec v })\subset F_L$ of faces, surrounding the vertex $\vec v$, see Fig.~\ref{fig:box}. The boxes are optimized by the algorithm in a specific order, namely in order of increasing value of $\Omega(\vec v)= \sum_m v_m L^m$. The set $E({\vec v })$ contains only those edges whose distance to $\vec v $, measured by the $l_\infty$ norm, is at most 1. Moreover, $E({\vec v })$ does not contain edges which are also part of the coarse-grained lattice: 
\begin{align} \nonumber
 E({ \vec v})  = &\{ e \in E_L\backslash E_L^{\text{cg}} \text{ s.t. } \forall \vec w\in e, |\vec v - \vec w |_\infty \leq 1 \} \ \ .
\end{align}
The set $F({\vec v})$ only contains those faces which contain an edge in $E({\vec v})$ {\em and} which contain only edges which are either an element of a set $E({\vec w})$ for which $\Omega(\vec w) \geq \Omega(\vec v)$ or which are an element of the coarse-grained lattice $E^{\text{cg}}_L$. The latter requirement prevents that moving errors out of a certain box corrupts a box which has already been cleared of errors. Formally one has
\begin{align}\nonumber
 &F({\vec v})  =\\ \nonumber
 &\left\{f\in F_L \text{ s.t. } 
 \begin{array}{l}
  \exists e\in E({ \vec v}) : e\subset f \text{ and } \\
  \forall e\subset f, e\in E^{\text{cg}}_L\cup \bigcup_{\Omega(\vec w) \geq \Omega(\vec v)}E({\vec w}) 
 \end{array}
\right\}\ \ .
\end{align}
Boxes contain at most 152 edges and 160 faces in the 4D hypercubic lattice and contain at most 650 edges and 2100 faces in the 5D hypercubic lattice. See Fig.~\ref{fig:box} for an illustration of an analogous box in the $L=5$ cubic code. The optimization for a box $\mathcal{B}(\vec v)$ is to find a $f_{\text{cg}}(\vec v)=\sum_{f \in F(\vec v)}  \alpha_f f$, $\alpha_f \in \{0,1\}$ which solves
\begin{align} \nonumber
\text{min} \ w\cdot f_{\text{cg}}(\vec v) \ \ \text{such that } \left.\ e_{\text{synd}}\right|_{E({\vec v})}= \left.\partial_2(f_{\text{cg}}(\vec v))\right|_{E(\vec v)}\ \ .
\end{align}
Here the boundary constraint, using $\partial_2$, uses mod 2 arithmetic and we use $|_{E({\vec v})}$ to denote the restriction to the space spanned by edges in $E(\vec v)$. This optimization over $O(1)$ variables can be recast into an integer program using slack variables, see e.g. page 8 in \cite{LAR:colorcode}. 
We believe and observe numerically that there always exists a $f_{\text{trial}}$, such that when restricted to the space spanned by faces in $F(\vec v)$, its boundary is equal to $\left.e_{\text{synd}}\right|_{E({\vec v})}$, although we do not prove this formally here.
After optimization of the box $\mathcal{B}({\vec v})$ the syndrome and the weight vector are updated to deal with the next box: $ e_{\text{synd}}\mapsto  e_{\text{synd}} + \partial_2 ( f_{\text{cg}}(\vec v))$, $w \mapsto (-1)^{f_{\text{cg}}(\vec v)}\cdot w$. The total correction of the RG step is eventually $f_{\text{cg}} = \sum_{\vec v} f_{\text{cg}}(\vec v)$.

\section{Results}
\label{sec:results}
We start by describing our performance metric for the decoders. For perfect measurements, after applying a correction using the RG decoder, one is guaranteed to be back in the code space. Correction is then successful if the product of errors and correction commutes with all logical operators. For faulty measurements one can perform a fixed amount, say $T-1$, error correction cycles after which one performs a single perfect measurement. This last measurement ensures that one can find a correction that maps back into the code space. Again correction is successful when the product of all errors and the correction on the qubits commutes with all logical operators. We use this method to assess the performance of the renormalization decoder with faulty measurements, setting the number of measurements $T$ equal to the system size $L$. Both methods give rise to a logical failure probability $\overline p_L$  (where the subscript $L$ refers to the code size). For the faulty measurement case, $\overline p_L$ should be interpreted as the failure probability within a time interval $T$ and could be normalized to obtain a failure probability per correction cycle.

Alternatively, for a single-shot decoder, one can perform a correction in each error correction cycle. After each such cycle errors potentially still remain, but if one is ``below'' threshold these errors should be primarily correctable. Thus if these remaining errors cannot be corrected by the same decoder using perfect measurements, the data is said to be corrupted. The memory time $T^{\text{mem}}_L$ is defined as the average number of error correction cycles before corruption. We use this method to assess the single-shot decoder. It can be related to a failure probability per cycle by assuming that $T_L^{\text{mem}} = \braket{t} = \sum_{t=0}^{\infty} t\overline p(1-\overline p)^t = \frac{1-\overline p}{\overline p}$. We report on thresholds as the crossing points between the curves $\bar p_L(p)$ or $ T^{\text{mem}}_L(p)$ for different $L$.

\begin{figure}[hbt]
 \begin{subfigure}[b]{.45\textwidth}
 \includegraphics[width = 1\textwidth]{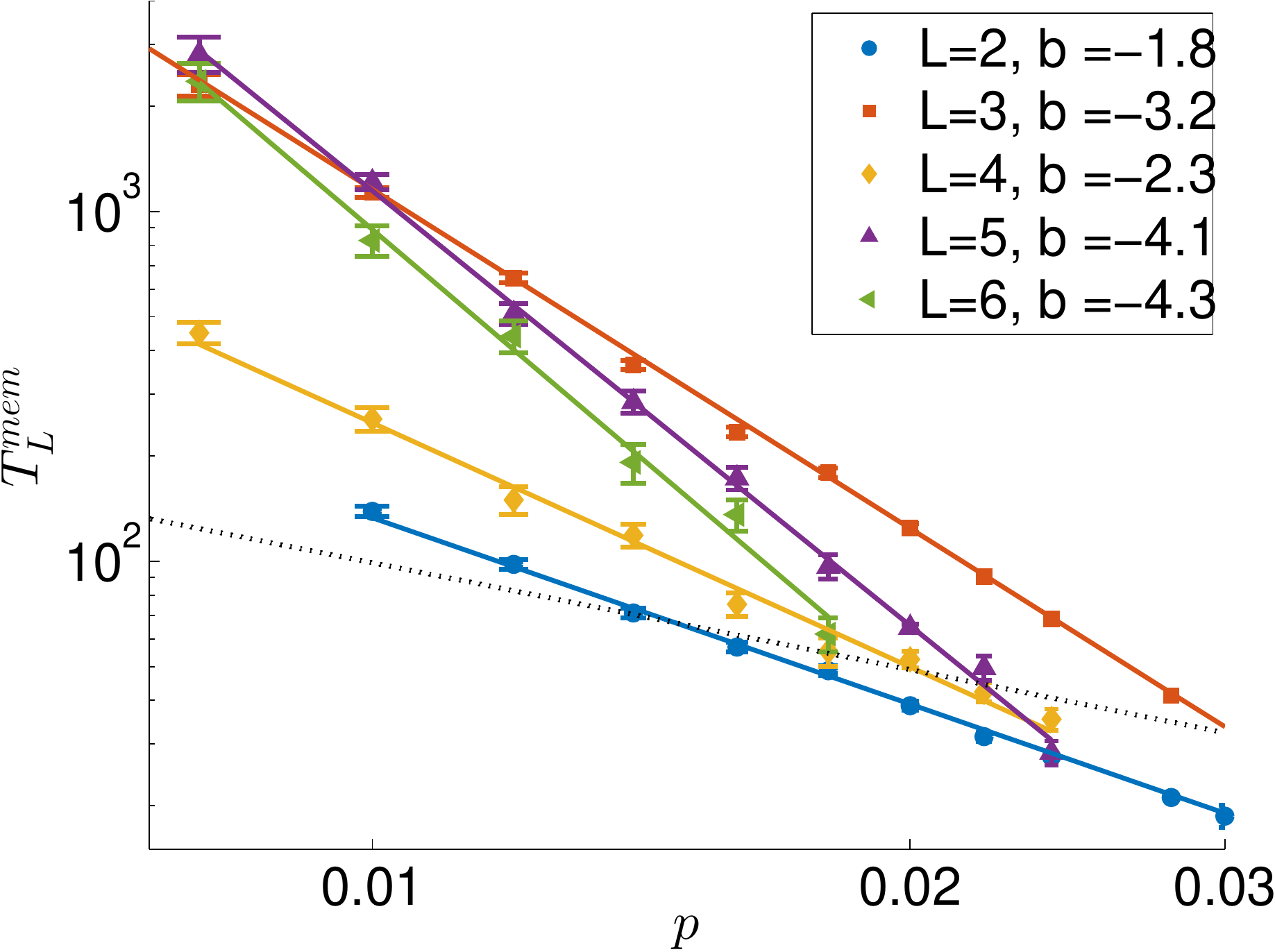}
 \caption{ $4D$}
 \label{fig:global_4D}
 \end{subfigure}
 
 \begin{subfigure}[b]{.45\textwidth}
 \includegraphics[width = 1\textwidth]{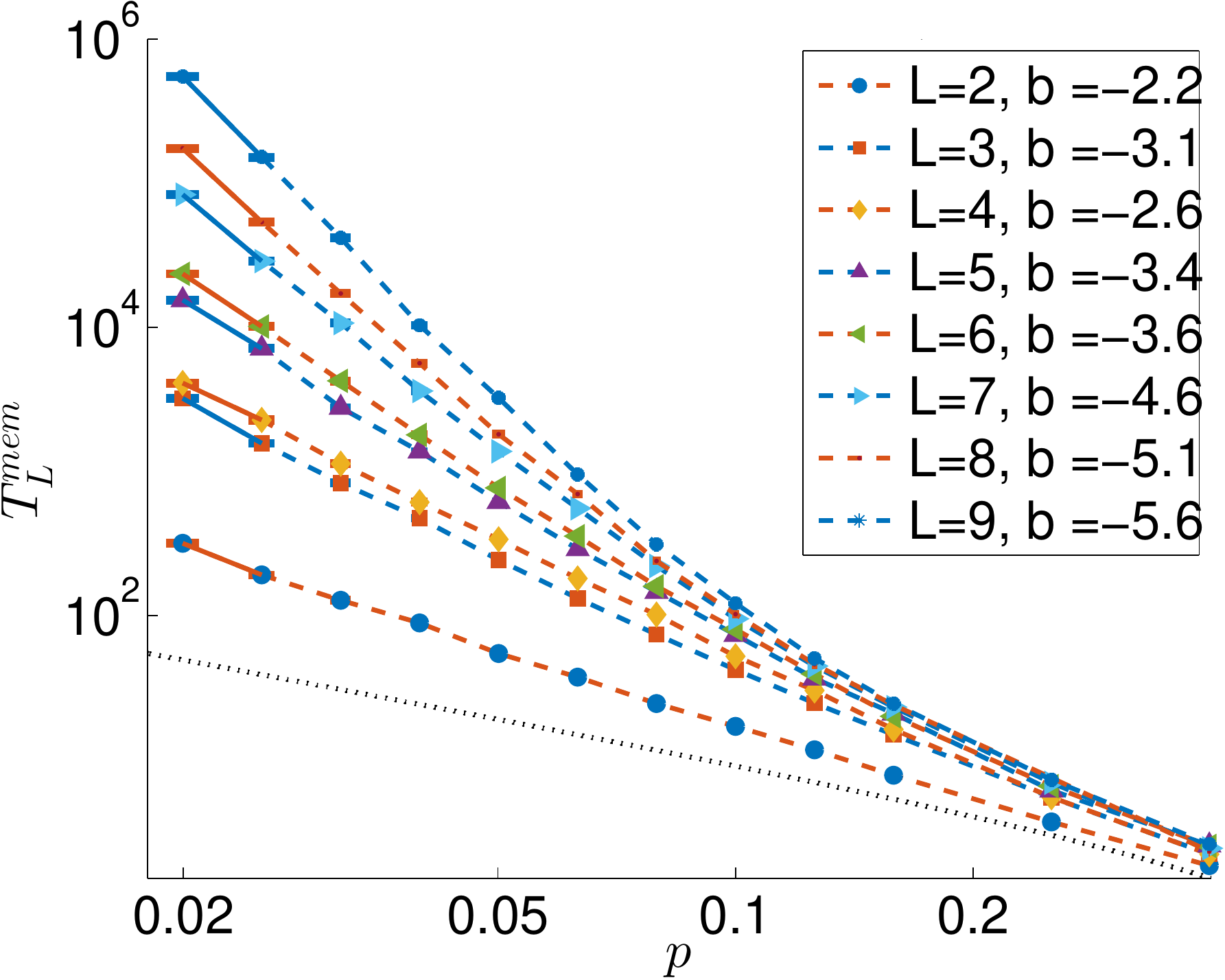}
 \caption{ $2D$}
 \label{fig:global_2D}
 \end{subfigure}
 \caption{ \label{fig:global}(color online) Memory time depending on the error strength for different system sizes $L$ with faulty measurements using the single-shot decoder. Panel (a): tesseract code. Panel (b): single-shot decoder applied to the two-dimensional version of the tesseract code corresponding to the 2D Ising model. Linear interpolating lines are $\overline p_L = ap^{b}$, with $ b $ given in the legend. Black dotted line gives the function $(1-p)/p$, the memory time of an un-encoded qubit, for reference.}
\end{figure}

\subsection{Performance of Single-shot Repair-Syndrome Decoder}
\label{sec:perf_ss}
To assess the effectiveness of the single-shot decoder which first corrects syndrome $ e_{\text{synd}}$ to form closed loops in 4D space, see Section~\ref{sec:ss}, we use a brute-force integer linear program to solve the second step, namely finding a $f_{\text{cor}}$ such that $\partial_2(f_{\text{cor}}) =  e_{\text{synd}} + e_{\text{cor}}$. In Fig.~\ref{fig:global_4D} we report on the memory time depending on the error probability $p$ for different system sizes. The largest code we consider has length $L=6$ and parameters [[5521,1,36]]. Although we cannot distinguish a clear threshold, we predict from this data that it is upper bounded by 2\% and hence lower than the threshold of the surface code for the same error model. For this reason we have not attempted to combine this single-shot decoder with a renormalization group decoder. 

\begin{figure}[bth]
\centering
 \includegraphics[width = 0.3\textwidth]{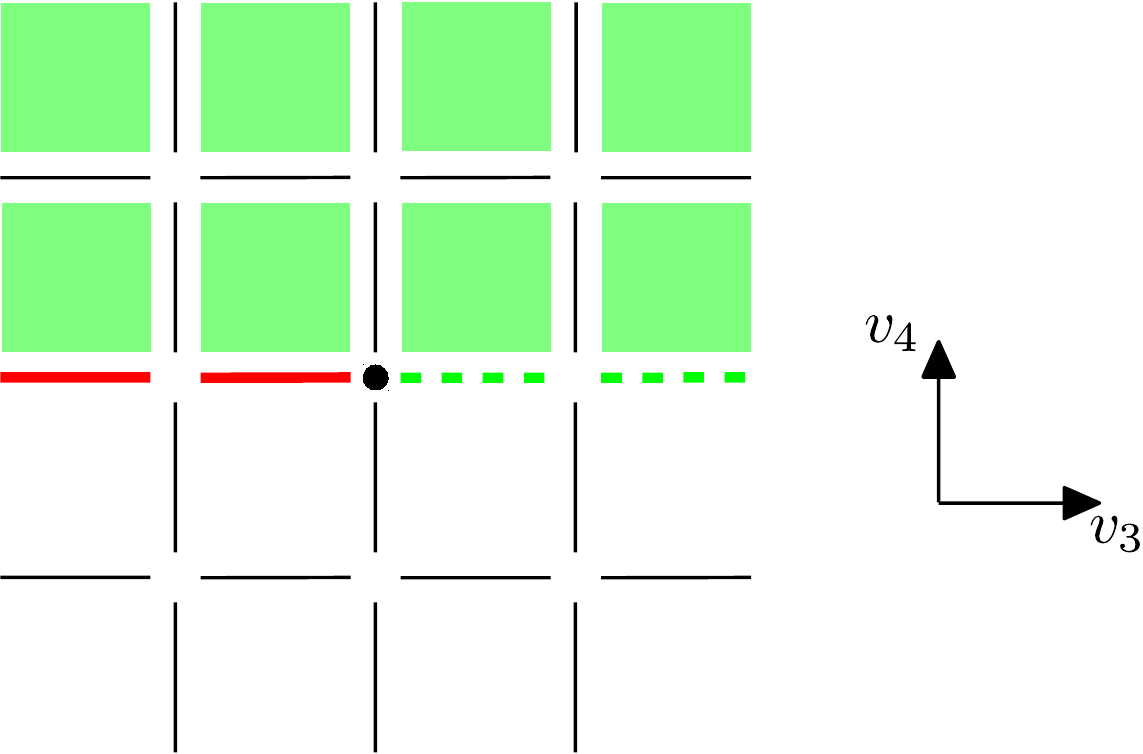}
 \caption{\label{fig:odd_even_effect}(color online) Illustration of how two measurement errors can lead to logical failure in the $L=4$ tesseract code. Depicted is a $(v_1,v_2) = {constant}$ cross-section. Red thick lines indicate the two measurement errors, the black dot indicate its end points $\partial_1(e_{\text{error} })$, which is also the endpoint of the measurement correction, depicted by two green dotted lines. (Note that there are actually four different ways to optimally correct this syndrome). The eight green squares indicate the qubit correction. Checking whether the memory is corrupted is done by a perfect measurement, giving rise to the boundary of the 8 flipped qubits, again being the two red and the two green lines. Since the algorithm of finding a minimal surface is not deterministic, finding a corresponding minimal surface could amount to flipping qubits corresponding to the lower 8 faces. The 16 qubits which are flipped in total form a logical operator and hence logical failure occurs.}
\end{figure}

Interestingly, we see that the memory time is worse for the $L=4$ code as compared to the $L=3$ code, not only quantitatively but also in its scaling with respect to $p$. This seems to be due to an odd-even effect. This effect also plays a role in a two-dimensional version of the tesseract code, which is obtained by setting $L_1=L_2=1$ and corresponds to the 2D Ising model. For errors which induce the surface-like logical operator of the 2D Ising model, one can repair the faulty syndrome using the minimum-weight matching algorithm and then pick the smallest of the two compatible surfaces. The data in Fig.~\ref{fig:global_2D} suggest a threshold in the $10-20\%$ range which is in fact comparable to the $17.2\%$ threshold lower-bound of the (non single shot) space-time RG decoding discussed in the next section. Similar to the tesseract code  we observe that the scaling of memory time with $p$ is again worse for $L=4$ as compared to $L=3$ and only slightly better for $L=6$ as compared to $L=5$. In Fig.~\ref{fig:odd_even_effect} we explain that, for a $L=4$ code, the memory can be corrupted within a single correction cycle with only two errors, which is not possible for a $L=3$ code.

\begin{figure*}[hbt!]
\begin{subfigure}[b]{.45\textwidth}
 \includegraphics[width = 1\textwidth]{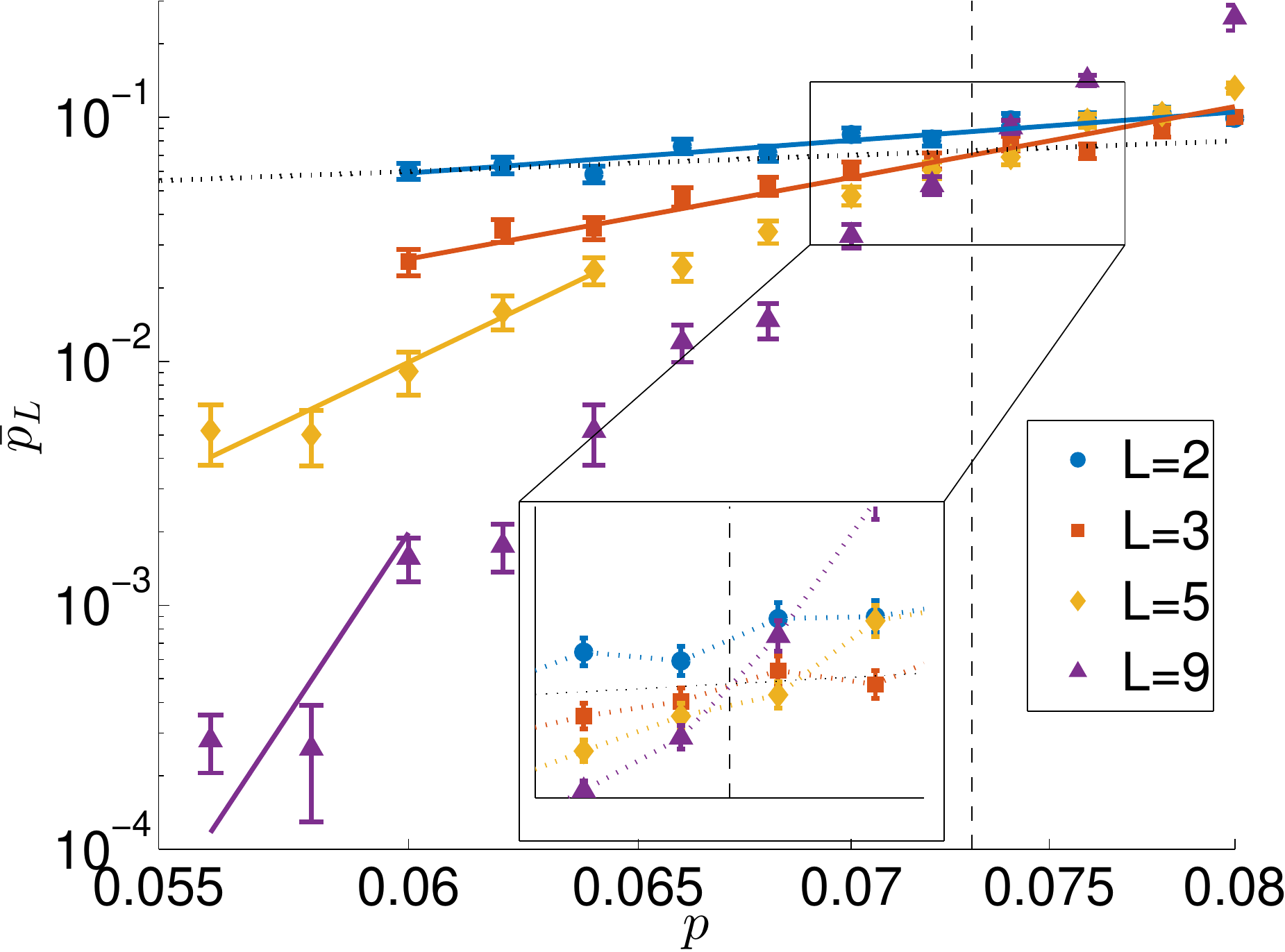}
 \caption{ $p_{\text{th}} = 7.3\pm0.1\% $}
 \label{fig:res_perf_meas}
 \end{subfigure}
 ~
 \begin{subfigure}[b]{.45\textwidth}
 \includegraphics[width = 1\textwidth]{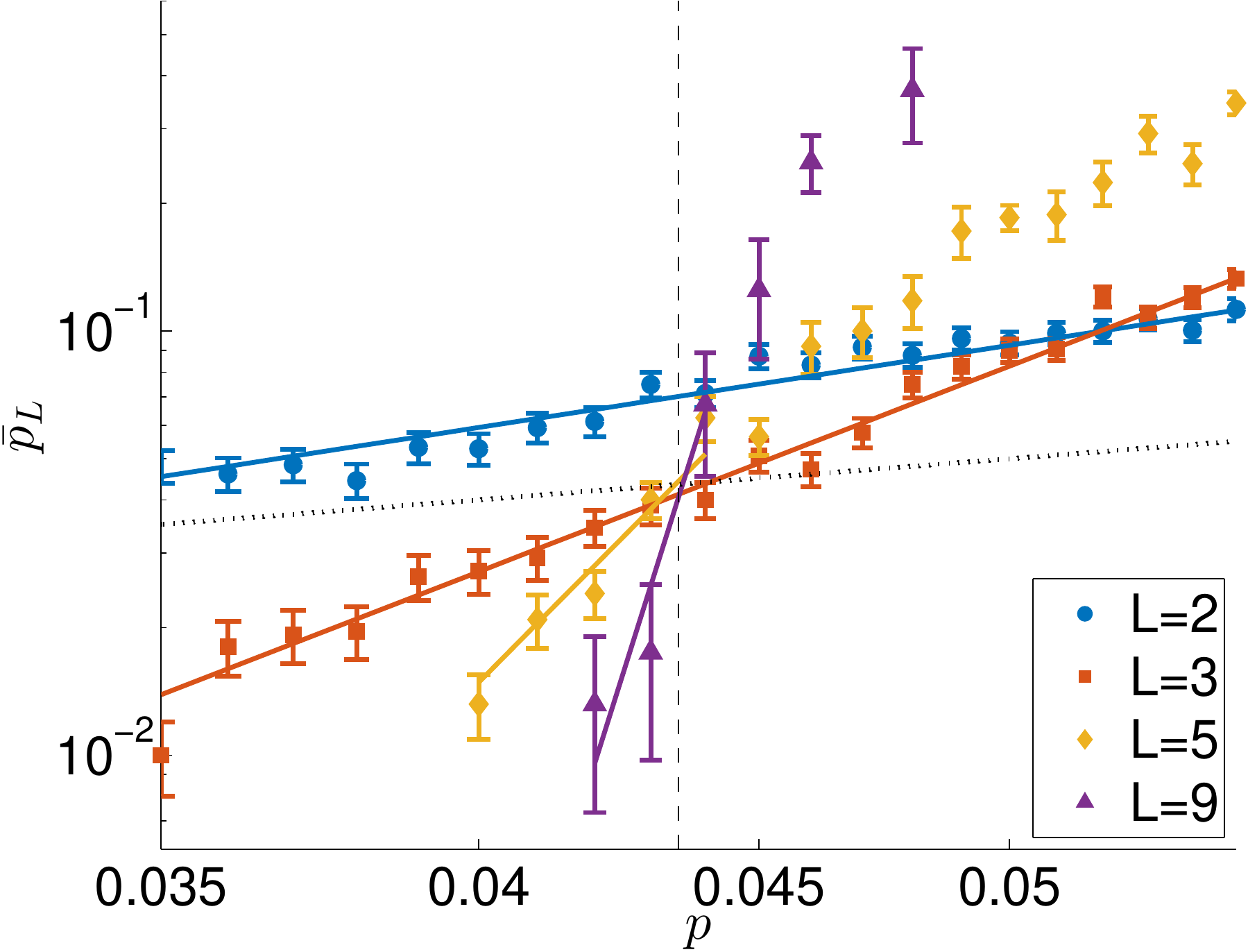}
 \caption{ $p_{\text{th}} = 4.35\pm0.1\% $}
 \label{fig:res_fem}
 \end{subfigure}
 
 \begin{subfigure}[b]{.45\textwidth}
 \includegraphics[width = 1\textwidth]{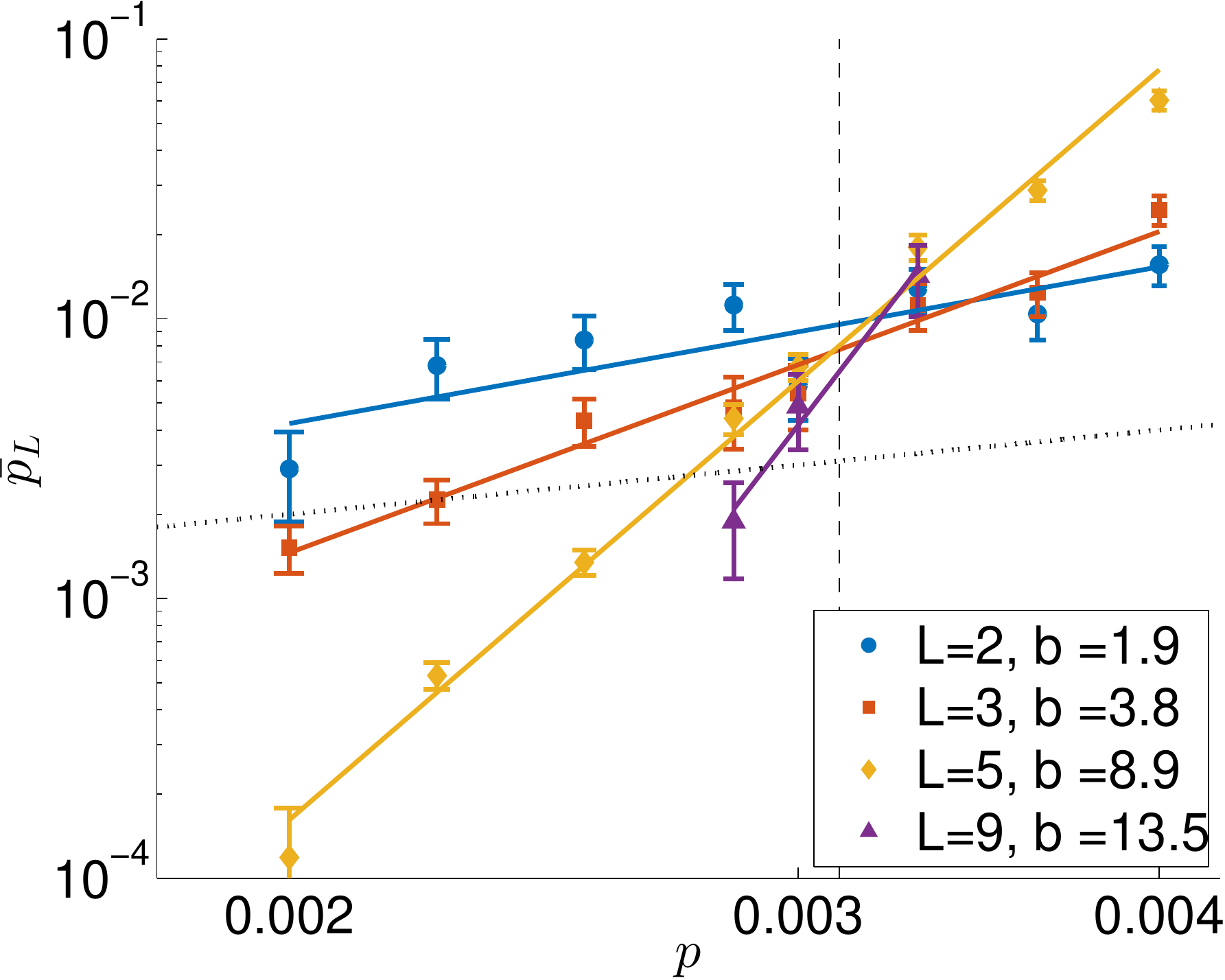}
 \caption{ $p_{\text{th}} = 0.31\pm0.01\% $}
 \label{fig:res_gb}
 \end{subfigure}
 ~
 \begin{subfigure}[b]{.45\textwidth}
 \includegraphics[width = 1\textwidth]{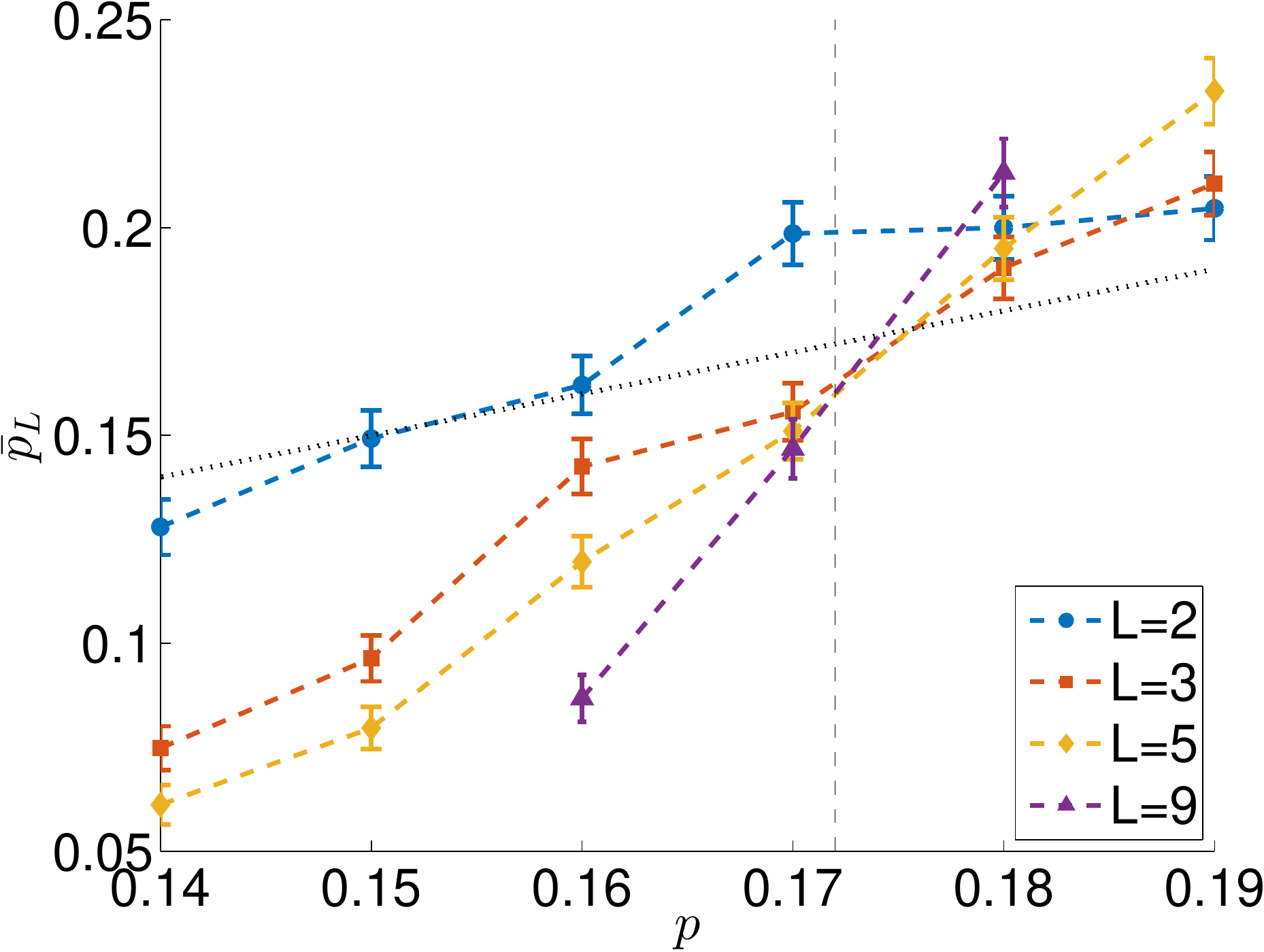}
 \caption{ $p_{\text{th}} = 17.2\pm1\% $}
 \label{fig:res_3d}
 \end{subfigure} 
 \caption{ \label{fig:res}(color online) Effective logical error $\bar p_L$ depending on error strength $p$ for different system sizes $L$ and different error models: panel a) phenomenological error with perfect syndrome measurement, panel b) phenomenological error with faulty syndrome measurement, panel c) gate-based errors and panel d) cubic code with phenomenological errors with perfect syndrome measurements. Linear interpolating lines are $\overline p_L = ap^{b}$, with $ b={\lceil \frac{L^2}{2} \rceil} $ for panel a) and b). Black dotted line gives the function $\overline p_L = p$, the error of an un-encoded qubit, for reference. Black vertical dashed lines indicate the location of the threshold, for reference.}
\end{figure*}

\subsection{Renormalization group decoder}
\label{sec:RG}
To assess the performance of the renormalization group decoder we study the crossing points of the curves $\overline p_L(p)$ for different system sizes $L\in\{2,3,5,9\}$. For all error models we observe that the crossing point between $\overline p_2$ and $\overline p_3$ occurs at a substantially higher probability $p$ as compared to the crossing points between the three curves of $\overline p_3$, $\overline p_5$ and $\overline p_9$. We attribute this to finite-size effects. Also note that decoding the $L=2$ tesseract code does not require coarse-graining. The three curves of $\overline p_3$, $\overline p_5$ and $\overline p_9$ do cross each other in a single point (within accuracy) and we report on this point as the threshold. Note that this number should be taken with caution due to the small number of system sizes and the lack of analytical proof for the existence of a threshold with this decoder.

For the phenomenological error model, with perfect measurements, we observe a threshold of $7.3\pm0.1\% $. This number is lower than the conjectured theoretical optimum using a maximum likelihood decoder ($11.003\%$ \cite{takeda}). The $\overline p_2$ and the $\overline p_3$ curve are consistent with a $ap^{\lceil \frac{d}{2} \rceil}$ behavior, where $d=L^2$ is the distance of the code. The $\overline p_5$ curve seems to follow this behavior only substantially below the threshold, that is, below $6.5\%$. We cannot confirm if the line $\overline p_9$ follows this behavior within the range of $p$ we considered.

For the phenomenological error model with faulty measurements, we observe a threshold of $4.35\pm0.1\% $. All data is consistent with the scaling $c_Lp^{\lceil \frac{d}{2} \rceil}$, suggesting that the threshold is unaltered when considering the error rate per QEC cycle $p^{\text{round}}_L = \frac{1}{2}(1-(1-2\overline p_L)^{1/L}) \approx \overline p_L/L$. 

When considering the gate-based error model, the threshold is substantially lower, namely we find $0.31\pm0.01\%$ for $Z$ errors. Moreover, the curves have a scaling which is worse than $p^{\lceil \frac{d}{2} \rceil}$.  This can be explained by propagation of errors through the quantum circuit. During a stabilizer measurement a single error could lead to up to three qubit errors (modulo the stabilizer). A  $p^{\lceil \frac{d}{6} \rceil}$ scaling is also not observed since it might only be valid for substantially lower values of $p$.

The decoder discussed in this paper can also be used to the decode the cubic code, having one surface-like logical operator and one line-like logical operator \cite{HZW:3D}. We found that for perfect syndrome measurement the surface-like logical has a threshold of $17.2\pm 1\%$, see Fig.~\ref{fig:res_3d}. The threshold of its line-like logical partner is expected to be the threshold of the surface code under faulty syndrome measurements, as the decoding problem of the two dimensional surface code with faulty measurement maps directly onto decoding the cubic code \cite{WHP:threshold,DKLP}.

\section{Conclusions}
\label{sec:con}
Although the tesseract code allows for a single-shot repair-syndrome decoding procedure, we find that this method is not competitive with earlier reported thresholds of the surface code. Even though a single-shot repair-syndrome decoder will have a threshold, the threshold is most likely not the same as the optimal threshold obtained by space-time decoding in 5D as we have argued in Section \ref{sec:ss} (see also Appendix \ref{A:HD}).
  
The renormalization group decoder that we introduce treats measurement and qubit errors on the same level. As far as we are aware, its threshold of $4.35\pm0.1\% $ is higher than any other threshold reported before for the same error model. The optimal threshold of the surface code (using a maximum likelihood decoder) is estimated to be $3.3\%$ by \cite{OAIM:3D} and estimated to be lower bounded by $2.9\%$ by \cite{WHP:threshold}. We have no reason to believe that the threshold for the four-dimensional toric code using the same decoder, will be different than the tesseract code.

Although the RG decoder leads to a threshold of the tesseract code which is higher than that of the surface code under a phenomenological error model with faulty measurements, it is still at least a factor of two lower than the theoretical optimum of $11.003\%$ \cite{takeda} and almost a factor of two lower than our own obtained threshold of 7.3\%, both corresponding to a phenomenological error model with perfect measurements. One could study how the threshold of the tesseract code behaves for different values of $p \neq q$, as was done in \cite{thesis:harrington} Chap. 4.5.4 for the toric code, to make a more thorough analysis of the effect of faulty measurements. 

Even though the RG decoder introduced in this paper focuses on the problem of finding a minimal surface given a boundary, its philosophy could equally well be applied to decoding any $(d_1,d_2)$-surface code and its performance could be compared with the RG decoder for the surface code of \cite{DP:fast, DP:RG} which uses message passing in addition to RG rescaling.

Ignoring the propagation of errors, the gate-based threshold can be upper bounded by the threshold obtained using the phenomenological error model in the following way. During a single QEC round a single qubit is acted on by eight CNOTs and two depolarizing error channels (during ancilla preparation and readout). Hence, the effective probability of an $X$ error on a qubit is $(8\cdot\frac{8}{15}+2\cdot\frac{2}{3})p=5.6p$ and thus the gate-based error threshold for our decoder should be no higher than $\frac{4.35}{5.6}\% \approx 0.7\%$. This suggests that redesigning the decoder to incorporate correlated errors, in analogy with \cite{FWH:error_prop}, might improve the threshold of $0.31\pm0.01\%$. 

Alternatively, one could use Shor error correction using cat state ancillas \cite{shor:fault_tollerant_ec} to minimize correlated errors. This would imply that measurement data would be less 
reliable and such scheme would still require eight rounds of CNOTs on each data qubit. Thus whether such an approach is beneficial is unclear at this moment, but a $p\neq q$ threshold curve might shed light on the matter. Another question is whether one can locally modify the code so that the qubit degree is lower while preserving or increasing the parity-check weight, for example by locally concatenating with the $[[4,2,2]]$-code.
The fact that $0.31\pm0.01\%$ is not competitive with the surface code might be purely be due to lack of decoder strength. Interestingly, using a decoder based on a neural network \cite{NXB:neural_networks} one obtains very similar thresholds for phenomenological error models, with and without measurement errors. To find out whether any decoder of the four-dimensional toric code, or tesseract code, is competitive with the surface code one could study the corresponding five-dimensional $\mathbb{Z}_2$-lattice gauge theory with quenched disorder (see e.g. \cite{CJR:lattice_gauge} for Monte Carlo studies without disorder).

\section{Acknowledgements}
KD was partially supported by the Excellence Initiative of the DFG. BMT, KD and NB acknowledge support through the ERC Consolidator Grant No. 682726.


\appendix

\subsection{Counting vertices, edges and faces}
\label{A:counting}
In this Appendix we will count the number of vertices, edges and faces of a tesseract code of size $L$. This can most easily be done by realizing that every cell is a product of intervals $o = \prod_i [a_i,b_i]$, with $b_i-a_i \leq 1$. The coordinates are restricted to:
\begin{align} \nonumber
a_i &\geq 0, &b_i&\leq L-1  &\forall i\in\{1,2\}\ \ ,\\ \nonumber
a_i &\leq L-1, &b_i&\geq 1  &\forall i\in\{3,4\}\ \ .
\end{align}
For vertices we have that $b_i=a_i$ and hence $|V_L| = L^2(L-1)^2$. For edges there is only one $i$ for which $b_i=a_i+1$. There are $(L-1)L(L-1)^2$ edges oriented in the $\vec a_1$ and $\vec a_2$ direction, and there are $L^3(L-1)$ edges oriented in the $\vec a_3$ and $\vec a_4$ direction, which in total gives $|E_L| = 2 L(L-1)^3 + 2 L^3(L-1)$. The cube set is of equal size by duality, see Appendix~\ref{sec:dual}.
Faces can be oriented in six different directions. The number of faces oriented in a certain direction again depends on that direction. It can be calculated in an analogous manner and results in:
\begin{align} \nonumber
\begin{array}{cc}
\text{orientation of faces}&\text{number of faces}\\\hline
 f_{\{1,2\}}(\vec v)&(L-1)^4 \\
 f_{\{1,3\}}(\vec v), f_{\{1,4\}}(\vec v), f_{\{2,3\}}(\vec v), f_{\{2,4\}}(\vec v)&L^2(L-1)^2 \\
 f_{\{3,4\}}(\vec v)&L^4
\end{array}
\end{align}
giving rise to a total of $(L-1)^4 + 4 L^2(L-1)^2 + 4L^2$ faces.

\subsection{Small codes}
\label{A:smallcodes}
In this Appendix we present the number of qubits and distance of small tesseract codes and compare them to surface codes and cubic codes of comparable distances. Following the argument in Appendix~\ref{A:counting}, the number of qubits for a tesseract code of size $L_1$ by $L_2$ by $L_3$ by $L_4$ is given by
\begin{align}\nonumber
 L_1L_2L_3L_4 + (L_1-1)(L_2-1)(L_3-1)(L_4-1) + \\ \nonumber
 [L_1(L_2-1)+(L_1-1)L_2][L_3(L_4-1)+(L_3-1)L_4] \ \ .
\end{align}
From Eqs.~\eqref{eq:logx} and \eqref{eq:logz} it can be understood that the minimal support of a logical $\overline X$ ($\overline Z$) operator is $L_3L_4$ (respectively $L_1L_2$). A tesseract code is clearly larger than a surface code for the same distance. A cubic code obtained by setting one of the dimensions to $1$, can be seen as a trade-off between the two. By choosing all lengths unequal to one can construct a variety of \textit{rectangular} codes of different sizes, see Table~\ref{tab:smallcodes}.

\begin{table}[htb]
\centering
 \begin{tabular}{r|cccc|cc}
  &$L_1$&$L_2$&$L_3$&$L_4$&$n$&$d$\\ \hline\hline
  tesseract&1&1&1&1&1&1\\
  codes&2&2&2&2&33&4\\
  $n\propto6d^2$&3&3&3&3&241&9\\
  &4&4&4&4&913&16\\ \hline
  surface&4&1&4&1&25&4\\
  codes&9&1&9&1&145&9\\
  $n\propto2d^2$&16&1&16&1&481&16\\ \hline
  rectangular&2&3&2&3&89&6\\
  4D&3&4&3&4&469&12\\
  &2&8&4&4&847&16\\
  \hline
  rectangular&4&1&2&2&28&4\\
  3D&6&1&2&3&71&6\\
  $n\propto3d^2$&9&1&3&3&177&9\\
  &12&1&3&4&331&12\\
  &16&1&4&4&616&16\\
 \end{tabular}
\caption{Number of physical qubits $n$ and distance $d$ of various small codes which encode a single qubit. All codes are obtained by varying the four linear dimensions $L_1$, $L_2$, $L_3$ and $L_4$ in the construction in Sections~\ref{sec:hom_des} and~\ref{sec:cel_com} under the constraint that $L_1L_2 = L_3L_4=d$. Note that the necessary choice $L_2=1$ for the cubic code makes the distance scale as $L_1$ which is a reflection of the fact that the logical $\overline X$ operator is line-like for the cubic code.}
\label{tab:smallcodes}
\end{table}

\subsection{Duality}
\label{sec:dual}
A feature that is understood about the 2D toric (or surface) code and the 4D toric code is that the cellular complexes on which these codes are based are self-dual. We can argue that the same feature holds for the tesseract code, i.e. one can show that the code is self-dual up to a rotation of the complex. This is a relevant conclusion for two reasons. When analyzing the performance of the code in terms of correcting independent $X$ and $Z$ errors, we only need to consider one of the two ($X$ or $Z$ errors). The duality can be used to perform a Hadamard gate transversely as in the surface code, and if required, the complex could be rotated back by code deformation around the boundaries (as was first done for the surface code in \cite{DKLP}).

The following duality transformation maps edges into cubes and faces into faces:
\begin{align} \nonumber
 e_{\{i\}}(\vec v)  \rightarrow e^*_{\{i\}}(\vec v) &= c_{{\rm All}\backslash i}(\vec v +\vec a_i ) \ \ , \\ \nonumber
 f_{\{i,j\}}(\vec v)  \rightarrow f^*_{\{i,j\}}(\vec v) &=f_{{\rm All}\backslash \{i,j\}}(\vec v +\vec a_i +\vec a_j )\ \ , \\ \nonumber
 c_{\{i,j,k\}}(\vec v) \rightarrow c^*_{\{i,j,k\}}(\vec v) &= e_{{\rm All} \backslash \{i,j,k\}}(\vec v +\vec a_i +\vec a_j +\vec a_k )\ \ ,
\end{align}
where ${\rm All}\backslash S$ uses ${\rm All}=\{1,2,3,4\}$. Conventionally, such duality mappings are also accompanied with an additional shift of half a lattice-spacing in all directions. The dual tesseract can also constructed by defining $U^* = [0,L]^2\times [1,L]^2$ and $B^*=\{\sum_i v_i\vec a_i\in U | v_1\in\{0,L\} \text{ or } v_2\in\{0,L\}\}$ and considering only those cells which are contained in $U^*$ but not fully contained in $B^*$. 

Duality states that
\begin{align} \nonumber
\left.\begin{array}{l}
o\subset U\\
o \not \subset B
\end{array}\right\}\Leftrightarrow\left\{\begin{array}{l}
o^*\subset U^*\\ 
o^*\not \subset B^*
\end{array}\right. \ \ .
\end{align}
This can be seen by using the interval representation $o = \prod_i [a_i,b_i]$ of cells, also used in Appendix~\ref{A:counting}. The dual cell $o^* = \prod_i [a^*_i,b^*_i]$ satisfies $a^*_i = b_i$ and $b^*_i=a_i+1$. The above inclusions of $o$ in the spaces $U$ and $B$ can be recast in terms of $a_i$ and $b_i$:
\begin{align} \nonumber
&\left.\begin{array}{lll}
a_i \geq 0, &b_i\leq L-1  &\forall i\in\{1,2\}\\
a_i \leq L-1, &b_i\geq 1  &\forall i\in\{3,4\}
\end{array}\right\}\Leftrightarrow\\ \nonumber
&\ \ \ \ \ \ \ \ \ \ \ \ \left\{\begin{array}{lll}
 a^*_i \leq L-1, & b^*_i\geq 1  &\forall i\in\{1,2\}\\
a^*_i \geq 1, & b^*_i\leq L  &\forall i\in\{3,4\}
\end{array}\right. \ \ ,
\label{eq:char_cel}
\end{align}
which can be straightforwardly checked. Also, the duality preserves inclusion in the sense that if $o_1 \subset o_2$ then the transformed cells obey $o_2^* \subset o_1^*$, showing that the code is self-dual (up to a rotation and a translation). 

\subsection{Logic}
\label{sec:logic}

Although a four-dimensional topological code is more challenging to implement, it potentially could allow for low-overhead or constant-depth constructions for non-Clifford gates, at least this is not precluded by the Bravyi-Koenig no-go theorem for low-dimensional codes \cite{BK:clifford, PY:ft}. CNOT gates can be performed by lattice code surgery in which logical $ZZ$ and $XX$ measurements are performed between two logical tesseract blocks at their boundaries and a logical ancilla tesseract block, in analogy with their implementation for a surface or cubic code, see e.g. \cite{CTV:review}. It is not clear whether mappings from color codes \cite{KYP:unfolding} would allow for logic beyond the Clifford group, in particular since the lowest-dimensional color code which has only surface-like logical operators and a transversal gate beyond the Clifford group is a six-dimensional color code \cite{bombin:selfcorrection}. 

In this Appendix we will prove that the tesseract code only allows for constant-depth constructions of logical gates which are elements of a restricted Clifford group. This proof makes an essential assumption on how blocks of tesseract code are glued together, in other words, what is the $O(1)$ neighborhood of a qubit in the code. It does not exclude all possible ways of using the tesseract code, e.g. it is not clear whether the ideas in \cite{KYP:unfolding} can give rise to non-Clifford constant-depth gates for this code. The restricted Clifford group consists of those Clifford operators mapping any Pauli $X$ to a product Pauli $X$ operators and similarly for $Z$ (Hadamard gates are excluded). An example is the CNOT gate.

Consider $n$ copies of the tesseract code with qubits labeled by a face $f$ and an index $i\leq n$. A local two-qubit gate acts on qubits corresponding to $(f,i)$ and $(f',j)$ such that the distance between the two faces $f$ and $f'$ is $O(1)$. A constant-depth gate consists of a finite-depth circuit of local two-qubit gates. Let the $\overline{X}_i$ and $\overline{Z}_i$ for $ i \leq n$ denote the logicals of each copy of the tesseract code. Consider any constant-depth gate keeping the code space invariant and let $U$ denote its restricted action to the code space. We will argue that (1) $V_i :=  U\overline{Z}_iU^\dagger$ commutes with any $\overline{Z}_j$ and that (2) $\overline{X}_jV_i\overline{X}_jV_i^\dagger=c_{ij}\mathbb{I}$ for some $c_{ij}$ for any pair $(i,j)$. Consider the representative of $\overline{Z}_j$ given in Eq.~\eqref{eq:logz} and consider an alternative ``moved-over" representative 
\begin{align}\nonumber
\overline{Z}_j^{\text{alt}} = \prod_{v_3,v_4=0}^{L-1} Z_{f_{\{3,4\}}((L-1)(\vec a_1+\vec a_2)+ v_3\vec a_3 +v_4 \vec a_4 )} \ \ .                                                                                                                       
\end{align}
We have $[\overline{Z}^{\text{alt}}_j, V_i]=0$ since $V_i$ has support on qubits corresponding to faces contained in the space $[0,d]^2\times[0,L]^2$ where $d$ is some $O(1)$ constant depending on the circuit depth and locality of the gates used. But the moved-over representatives $\overline{Z}_j^{\text{alt}}$ have support only on qubits corresponding to faces contained in $\{L-1\}^2\times[0,L]^2$ and hence all commute with $V_i$. 
Similarly, consider representatives for $\overline{X}_j$ given in Eq.~\eqref{eq:logx}, having support on qubits corresponding to faces contained in the space $[0,L-1]^2\times[0,1]^2$. The overlap of the support of $V_i$ and $\overline{X}_j$ is restricted to $O(d^2) = O(1)$ qubits and hence, following arguments in \cite{BK:clifford} we have that $\overline{X}_jV_i\overline{X}_j = c_{ij}V_i$. Moreover, due to Hermiticity of $V_i$, we have that $c_{ij}\in\{-1,1\}$.

From the commutation between $V_i$ and $\overline{Z}_j$ and since products of $\overline{Z}_j$ form a complete set of commuting logical observables, it follows that $V_i$ can be written as sums of products of $\overline{Z}_j$ or in other words $V_i$ is necessarily ``diagonal in the $Z$-basis'':
\begin{align}\nonumber
 V_i = \prod_{j=1}^n (\beta^0_j\mathbb{I} + \beta^1_j\overline{Z}_j) \ \ .
\end{align}
Since $\overline{X}_jV_i\overline{X}_j = \pm V_i$ we have that either $\beta^0_j = 0$ or $\beta^1_j = 0$ for any $j$. Hence $V_i$ can be written as a product of $\overline{Z}_j$ showing that $U$ is an element of the restricted Clifford group.

\subsection{CNOT ordering}
\label{A:cnot}
In this section we will go into details of the CNOT ordering used to measure the stabilizers of the tesseract code as described in Section~\ref{sec:gb}. Consider any two stabilizers $S^{X}_e$ and $S^{Z}_c$ having overlapping action on two qubits labeled by the faces $f_1$ and $f_2$. Consider the four CNOT operators acting between these qubits, we label them by the tuples $(e,f_1)$, $(c,f_2)$, $(e,f_2)$ and $(c,f_2)$. In order for the stabilizers to be measured correctly, either the CNOT of $(e,f_1)$ is performed before the CNOT of $(c,f_1)$ \textit{and then} $(e,f_2)$ should be done before $(c,f_2)$, {\em or} $(c,f_1)$ is performed before $(e,f_1)$ and then $(c,f_2)$ should be performed before $(e,f_2)$ as well. We show that the CNOT schedule as described in the main text has indeed this property for any pair of faces and overlapping $X$- and $Z$-stabilizers. This implies that one can fully interleave the circuits for $X$- and $Z$-stabilizer measurement, leading to 8 CNOT rounds. 

Let the cube corresponding to the $Z$-stabilizer be labeled as $c_{\{i,j,k\}}(\vec w)$ and let the edge corresponding the $X$-stabilizer be labeled as $e_{\{l\}}(\vec v)$. In order for these stabilizers to have overlapping support one needs at least $l \in \{i,j,k\}$. We assume without loss of generality that $l=i$ and $j<k$ in which case overlapping stabilizers obey $\vec v=\vec w+s_j\vec a_j+s_k\vec a_k$ for $s_j,s_k\in\{0,1\}$. The two faces corresponding to the qubits on which both stabilizers have support are then labeled by $f_1  = f_{\{i,j\}}(\vec w+s_k\vec a_k)$ and  $f_2 = f_{\{i,k\}}(\vec w+s_j\vec a_j)$. See Fig.~\ref{fig:cnot} for an example in which $(i,j,k) = (3,1,2)$ and $(s_j,s_k) = (1,0)$.

\begin{figure}[tb]
\centering
 \caption{\label{fig:cnot}(color online) Example of the CNOT labeling for the case of $(i,j,k) = (3,1,2)$ and $(s_j,s_k) = (1,0)$. Gray squares represent faces/qubits $f_i$ and the thick blue line represents the edge/$X$-stabilizer $e$. The direction labeling the CNOT $(e,f_1)$ is $\vec a_1$ and the CNOT $(e,f_2)$ is $-\vec a_2$.}
\end{figure}

As explained in the main text, the 8 CNOT rounds are labeled by a direction $\vec d  = (-1)^n\vec{a_s}$, $n\in\{0,1\}$, $s\in\{1,2,3,4\}$ from qubit to ancilla. The ordering of CNOTs in terms of this direction  is $[-\vec a_1 , -\vec a_2 ,-\vec a_3 ,-\vec a_4 ,\vec a_4 ,\vec a_3 ,\vec a_2 ,\vec a_1 ]$. 
During a single round, a CNOT gate is applied between qubit $f_{\{i,j\}}(\vec v)$ and a cube ancilla at $c_{\{i,j,s\}}(\vec v +n\vec a_s)$ or an edge ancilla at $e_{\{i,j\}\backslash s}(\vec v +(1-n)\vec a_s)$, depending on whether $s\in\{i,j\}$. From this we can infer the direction label of the rounds in which the four CNOTs under consideration are applied, see Table.~\ref{tab:cnot}: we see that the ordering of the four CNOTs under consideration depends on the two labels $s_j$ and $s_k$. The ordering is such that either the $S^{X}_e$ CNOTs are performed before the $S^{Z}_c$ CNOTs or vice versa.

\begin{table}[htb]
\centering
 \begin{tabular}{cc|cccc}
 &&\multicolumn{4}{c}{$(s_j,s_k)$} \\
 CNOT&direction $\vec d$&(0,0)&(0,1)&(1,0)&(1,1)\\ \hline
 $(e,f_1)$&$-(-1)^{s_j}\vec a_j$&1&1&4&4\\
 $(e,f_2)$&$-(-1)^{s_k}\vec a_k$&2&3&2&3\\
 $(c,f_1)$&$(-1)^{s_k}\vec a_k$&3&2&3&2\\
 $(c,f_2)$&$(-1)^{s_j}\vec a_j$&4&4&1&1
 \end{tabular}
 \caption{ \label{tab:cnot}Temperal ordering (e.g. $1,2,\ldots$) of CNOTs associated with $(e,f)$ and $(c,f)$ depending on their relative locations given by $s_j$ and $s_k$. One observes that the ordering obeys the desired property.}
 \end{table}

 \subsection{Higher-dimensional Surface Codes and Equivalence Between Minimum-Weight Decoding Problems}
\label{A:HD}
In this section we generalize the family of surface codes to a $D$-dimensional hypercubic lattice for any dimension $D=d_1+d_2$. Here $d_1$ is the number of directions in which one has a ``smooth'' boundary and $d_2$ is the number of directions in which one has a ``rough'' boundary. These $(d_1,d_2)$-surface codes are again defined over cellular complexes of spaces namely the spaces
\begin{align} \nonumber
 U &= \prod_{i=1}^{d_1}[0,L_i-1]\times\prod_{i=d_1+1}^{D}[0,L_i]\ \ ,\\ \nonumber
 B &= \left\{\vec v\in U\ \  \text{s.t}\ \ \exists i>d_1 , v_i \in\{0,L_i\} \right\} \ \ .
\end{align}
Consider cells $o_I(\bf v)$ labeled by vertices $\vec v = \sum_{i=1}^D v_i\vec a_i$, with unit vectors $\vec a_i$ and integer coefficients $v_i$, defined as,
\begin{align} \nonumber
 o_I(\vec v) = \left\{\vec v+\sum_{k\in I} s_k \vec a_k \ \ |\ \ s_k \in [0,1] \right\}.
\end{align}
The set $I$ contains the orientation and its cardinality $|I|$ equals the dimensionality of the cell $o_I(\vec v)$. Qubits are defined on $d_2$-cells which are contained in $U$ but not contained in $B$. $X$- and $Z$-stabilizers are defined for $d_2-1$-cells and $d_2+1$-cells, respectively, which are contained in $U$, but not in $B$:
\begin{align} \nonumber
 S^{X}_o := \prod_{o':o\subset o'} X_{o'}, \ \  S^{Z}_o := \prod_{o':o'\subset o} Z_{o'}\ \ .
\end{align}
The number of encoded qubits is given by $\text{dim}(H_{d_2}(U,B)) = 1$. All other homology groups are trivial. The support of the logical $\overline{X}$ operator is $d_1$-dimensional and the support of the logical $\overline{Z}$ operator is $d_2$-dimensional. Explicitly:
\begin{align} \nonumber
 \overline{X} &= \prod_{v_1=0}^{L_1-1}\dots\prod_{v_{d_1}=0}^{L_{d_1}-1} X_{f_{\{d_1+1,\dots,D\}}(\sum_{i=1}^{d_1}v_i\vec a_i)} 
\ \ \text{and}\\ \nonumber
 \overline{Z} &= \prod_{v_{d_1}=0}^{L_{d_1}-1}\dots\prod_{v_{D}=0}^{L_{D}-1} Z_{f_{\{d_1+1,\dots,D\}}(\sum_{i=d_1+1}^{D}v_i\vec a_i)} \ \ .
\end{align}

Under a duality transformation of the cell complex (as in Section~\ref{sec:dual}), the code is transformed to a $(D-d_1, D-d_2)$-surface code, hence for $d_1=d_2=D/2$ the construction is self-dual. If one treats the $X$-stabilizers as the gauge symmetry of a Hamiltonian constructed itself from only $Z$-stabilizers, then these Hamiltonians will be Ising gauge models (without magnetic fields) as defined by Wegner \cite{wegner:ising}. 

Consider now the decoding problem for $Z$ errors for a $(d_1,d_2)$-surface code in the phenomenological error model described in Section~\ref{sec:phen} with syndrome as well as qubit errors, both with probability $p$. Repeated faulty syndrome measurements ($T$ times, labeled by $t\in\{0,1,..T-1\}$, with the last round being perfect) of all $X$-stabilizers of a $(d_1,d_2)$-surface code can be interpreted as forming the boundary of a $d_2$-dimensional error surface. The minimum-weight decoding problem is then to find a surface of minimal area given this boundary. 

More precisely, let $U_{\text{ST}}= U\times [0,T-1]$ and $B_{\text{ST}} = B\times[0,T-1]$, so that the time-boundary is `smooth' (not allowing for any hypersurface to attach). The $d_2$-cell $\tilde o_I(\vec v+t\vec a_{D+1})$ can either represent a measurement error if $D+1 \in  I$ (oriented in the time direction) or a qubit error if $D+1 \not\in  I$ (oriented in the spatial direction). In the former case it corresponds to a measurement error of the $X$-stabilizer corresponding to the $d_2-1$-cell $o_{I\backslash D+1}(\vec v)$ at round $t$. In the latter case it represents an error of the qubit corresponding to the $d_2$-cell $o_I(\vec v)$ occurring between measurements at times $t$ and $t-1$. 


The syndrome itself consists of $d_2-1$-cells $\tilde o_I(\vec v+t\vec a_{D+1})$ which are either oriented in the time direction ($D+1 \in  I$) or not. 
In the latter case it signifies a change in outcome of the faulty measurement of the stabilizer $o_I(\vec v)$, between rounds $t$ and $t-1$. In the former case it signifies a violation of the linear dependency relation
\begin{align} \nonumber
\prod_{o':o\subset o'} S^X_{o' } = \mathbb{I} \ \ ,
\end{align}
labeled by the $d_2-2$-cell $o=o_{I\backslash D+1}(\vec v) $ during measurement round $t$ (when $d_2<2$ we do not have such dependency). 
The problem of minimum-weight space-time decoding for the code given by $U$ with boundary $B$ can thus be stated as the minimum-weight decoding problem for a surface code associated with the complex $U_{\text{ST}}$ with boundary $B_{\text{ST}}$. The qubits of the higher-dimensional code are defined for each $d_2$-cell $\tilde o_I(\vec v+t\vec a_{D+1}) \in U_{\text{ST}}$ but not contained in $B_{\text{ST}}$, while the $X$-stabilizers are associated with the $d_2-1$-cells $\tilde o_I(\vec v+t\vec a_{D+1}) \in U_{\text{ST}}$ but not contained in $B_{\text{ST}}$. 
The logical $\overline{Z}$ operator of the higher-dimensional code has the same minimum weight as the logical $\overline{Z}$ of the lower-dimensional code: deformation in the time-direction cannot minimize the weight since the logical $\overline{Z}$ is stretched between the rough boundaries.
We conclude that the minimum-weight space-time decoding problem of a $(d_1,d_2)$-surface code (faulty syndrome measurements) can be cast as the problem of minimum-weight decoding in the $(d_1+1,d_2)$-surface code (perfect syndrome measurements). This mapping is not only expected to hold for minimum-weight decoding but could be straightforwardly extended to maximum-likelihood decoding as analyzed in \cite{DKLP, WHP:threshold}. It is for this reason that identical threshold (upper-bounds) in Table~\ref{tab:thresholds} are stated for codes related by this $D \rightarrow D+1$ mapping.
Similarly, for $X$ errors, the mapping identifies faulty syndrome decoding of a $(d_1,d_2)$-surface code with perfect decoding of a $(d_1,d_2+1)$-surface code (by first going to the dual complex in which rough and smooth boundaries are interchanged, then extending this in the time-direction with a smooth boundary to a space-time complex, and taking the dual again).

For sufficiently high dimensions $d_2 \geq 2$ one can consider single-shot decoding as described in Section \ref{sec:ss}. The syndrome consisting of $d_2-1$-cells is a closed hypersurface since it is itself the boundary of qubit and measurement errors on the $d_2$-cells. In single-shot decoding one considers a single time-slice $t$ and thus we project out, for a given $t$, any of the $d_2-1$ cells $\tilde o_I(\vec v+t\vec a_{D+1})$ which are oriented in the time direction, as well as any $d_2-1$ cell at coordinates $\vec v + t' a_{D+1}$ with $t' \neq t$. This leaves a $d_2-1$-dimensional hypersurface $e_{\text{synd}}$ in $D$ dimensions with a $d_2-2$-dimensional boundary. Repairing the syndrome can be done by finding a minimal $d_2-1$-dimensional hypersurface $e_{\text{cor}}$ given this boundary, where $\partial_{d_2-1}(e_{\text{synd}})=\partial_{d_2-1}(e_{\text{cor}})$. 
This step in decoding is effective when the repaired syndrome $e_{\text{cor}}+e_{\text{synd}}$ is the boundary of a set of correctable qubit errors, so that upon applying these corrections, the next round of single-shot decoding receives incoming errors according to some ``effectively local'' error model. One expects that precisely when the repaired syndrome starts to become homologically non-trivial (with boundaries of the lattice appropriately chosen), that is, when logical failure starts to happens in minimum-weight decoding of $Z$ errors of a $(d_1+1, d_2-1)$-dimensional surface code (i.e. for its $d_2-1$-dimensional logical operator), that the repaired syndrome may not lead to correctable qubit errors. Hence we conjecture that for, say, the tesseract code $(d_1=2,d_2=2)$ with faulty measurements, the threshold for single-shot decoding is bounded by the minimum of the thresholds for perfect measurements of the line-like logical in 4D ($(3,1)$-surface code) and the surface-like logical in 4D ($(2,2)$-surface code). Space-time decoding for the tesseract code corresponds to perfect measurement decoding of the surface-like logical in 5D ($(3,2)$-surface code), hence one expects that optimal thresholds of single-shot versus space-time decoding do not coincide.


\end{document}